\documentclass{article}
\usepackage[dvips]{graphicx}
\topmargin=-1cm\oddsidemargin=0truecm\evensidemargin=0truecm\textheight=24.0cm\textwidth=16.5cm\usepackage{color}
\def\vol#1#2#3{{\bf {#1}} ({#2}) {#3}}\def\NP{Nucl.~Phys. }\def\PL{Phys.~Lett. }\def\PR{Phys.~Rev. }\def\PRP{Phys.~Rep. }\def\PTP{Prog.~Theor.~Phys. }\def\IJMP{Int.~J.~Mod.~Phys. }

\def\no{\nonumber}

\def\2tvec#1#2{
\left(
\begin{array}{c}
#1  \\
#2  \\   
\end{array}
\right)}

\def\mat2#1#2#3#4{
\left(
\begin{array}{cc}
#1 & #2 \\
#3 & #4 \\
\end{array}
\right)
}

\def\Mat3#1#2#3#4#5#6#7#8#9{
\left(
\begin{array}{ccc}
#1 & #2 & #3 \\
#4 & #5 & #6 \\
#7 & #8 & #9 \\
\end{array}
\right)
}

\def\3tvec#1#2#3{
\left(
\begin{array}{c}
#1  \\
#2  \\   
#3  \\
\end{array}
\right)}

\def\4tvec#1#2#3#4{
\left(
\begin{array}{c}
#1  \\
#2  \\   
#3  \\
#4  \\
\end{array}
\right)}

\def\L{\left}
\def\R{\right}

\def\pl{\partial}

\def\hbar{\hspace{1mm}\bar{}\hspace{-1mm}h}

\def\eqn#1{
\begin{eqnarray}
#1
\end{eqnarray}
}

\begin{document}

\begin{titlepage}
\begin{flushright}
KANAZAWA-11-11, KIAS-P11037
\end{flushright}
\begin{center}

\vspace{1cm}
{\Large\bf Two Component Dark Matters\\
 in\\ $S_4\times Z_2$ Flavor Symmetric Extra U(1) Model}\\
 \vspace{0.5cm}
 Yasuhiro Daikoku$^{a}$\footnote{yasu\_daikoku@yahoo.co.jp},
Hiroshi Okada$^{b,c}$\footnote{hokada@kias.re.kr}
 and
Takashi Toma$^{a,d}$\footnote{t-toma@hep.s.kanazawa-u.ac.jp}
\vspace{5mm}

 {\it 
$^a${ Institute for Theoretical Physics, Kanazawa University, Kanazawa 920-1192, Japan} \\ 
 \vspace{1mm} 
$^b${ Centre for Theoretical Physics, The British University in Egypt},\\
 El Sherouk City, Postal No, 11837, P.O. Box 43, Egypt\\ \vspace{1mm} 
 $^c${School of Physics, KIAS, Seoul 130-722, Korea}\\ \vspace{1mm}
$^d${ Max-Planck-Institut f\"ur Kernphysik, Postfach 103980, 69029 Heidelberg, Germany}
}
  
  \vspace{8mm}

\begin{abstract}
We study cosmic-ray anomaly observed by PAMELA based on 
$E_6$ inspired extra U(1) model with $S_4\times Z_2$ flavor symmetry.
In our model, the lightest flavon has very long lifetime of
${\cal O}(10^{18})$ second which is longer than the age of the universe, but not long
enough to explain the PAMELA result $\sim{\cal O}(10^{26})$ sec. 
Such a situation could be avoidable by considering 
that the flavon is not the dominant component of dark matters and the dominant one is 
the lightest neutralino.
With appropriate parameter set, density parameter of dark matter
and over-abundance of positron flux in cosmic-ray are realized at the same time.
There is interesting correlation between spectrum of positron flux and $V_{MNS}$.
No excess of anti-proton in cosmic-ray suggests  that 
sfermions are heavier than 4 TeV and the masses of the light Higgs bosons are degenerated.
\end{abstract}

\end{center}
\end{titlepage}
\setcounter{footnote}{0}

\newpage

\section{Introduction}

Standard Model (SM) is successful theory of gauge interactions,
however Higgs sector is not examined well.
Therefore mass matrices of leptons and quarks are not well understood.
Many unsolved puzzles of SM are left in these sectors; that is, {\it e.g.},
why is the structure of mixing matrix of leptons (Maki-Nakagawa-Sakata matrix, $V_{MNS}$)  very different from that of the mixing matrix of quarks (Cabibbo-Kobayashi-Maskawa matrix, $V_{CKM}$),
especially why is the mixing angle $\theta_{23}$ maximal?
Why is neutrino mass far smaller than those of other fermions?
Why do generations exist?

We also find a problem in cosmology.
In modern cosmology, the existence of the dark matter is clear.
Recent cosmic-ray observation of PAMELA suggests that 
the dark matter decays mainly into leptons with very long lifetime \cite{pamela}\cite{positron}\cite{Ibarra:2008jk}.
Such a particle is not included in SM.

Separately from these puzzles, there is hierarchy problem 
why electroweak scale is much smaller than Planck scale.
One of the solutions is to introduce supersymmetry (SUSY) \cite{SUSY}.
However minimal supersymmetric standard model (MSSM) does not satisfy the solution,
because we must fine-tune $\mu$-parameter in superpotential of MSSM, which is
much smaller than Planck scale in order to realize appropriate electroweak symmetry breaking. This is called $\mu$-problem.

Another problem of MSSM is proton stability.
The R-parity forbids baryon number violating trilinear terms in
superptential, however does not forbid quartic terms like $E^cU^cU^cD^c,LQQQ$.
Such interactions reduce the lifetime of proton to unacceptable level \cite{proton-decay}.
Therefore the R-parity does not help the explanation of proton stability.
The problem of proton lifetime of supersymmetric model is one of the most essential point
in understanding generation structure.

With the motivation to solve flavor puzzles and hierarchy problem, we introduce new three symmetries.
At first, we introduce non-Abelian discrete flavor symmetry $S_4\times Z_2$,
in order to explain that the mixing angle $\theta_{23}$ is maximal \cite{kubo}\cite{s4e6}\cite{Kajiyama:2006ww}\cite{review}\cite{s4}.
Because $V_{CKM}$ and $V_{MNS}$ are very different, it is expected that
the representations of quarks and leptons are also different.
Next, we introduce $U(1)_X$ gauge symmetry which forbids $\mu$-term \cite{mu-problem}.
Then, several new superfields must be introduced
due to gauge anomaly cancellation condition;
those are extra Higgs $(H^U,H^D)$, singlet Higgs $S$ and exotic quarks $(g,g^c)$.
The extra Higgs bosons couple only to leptons,
which induce the difference between $V_{CKM}$ and $V_{MNS}$.
Moreover, the existence of exotic quarks is important to understand the meaning of generations.
Finally we introduce $U(1)_Z$ gauge symmetry.
Due to the anomaly cancellation condition, 
right-handed neutrino (RHN) superfield $N^c$ is introduced,
then the smallness of neutrino mass is realized by seesaw mechanism.
The two new $U(1)$ gauge symmetries and 
standard model gauge symmetry $G_{SM}=SU(3)_c\times SU(2)_W\times U(1)_Y$
can be embedded in $E_6$ as $G_{SM}\times U(1)_X\times U(1)_Z\subset E_6$,
then MSSM and new superfields consist {\bf 27} of $E_6$ representation.
With appropriate assignment of superfields under the flavor symmetry,
the stability of proton is realized, which plays the most important role in the flavor symmetry. Thus we can understand that the generation structure is the new system to stabilize proton \cite{s4e6}.

The new symmetries which are introduced above may also solve dark matter problems.
As three $U(1)$ gauge symmetries include R-parity, lightest
supersymmetric particle (LSP) is a candidate for dark matter. 
The positron flux observed by PAMELA is produced by the field
which induces RHN mass and decays into leptons \cite{u1-pamela}.
In this paper, we show our model is consistent with experimental results of dark matter.
At first, we define our model in section 2. The estimations of relic abundance of dark matter
and positron flux are given in section 3. Finally, we give conclusion of our analysis in section 4.

\section{$S_4\times Z_2$ flavor symmetric extra U(1) model}

\subsection{Gauge symmetry}

\begin{table}[htbp]
\begin{center}
\begin{tabular}{|c|c|c|c|c|c|c|c|c|c|c|c||c|c|}
\hline
         &$Q$ &$U^c$    &$E^c$&$D^c$    &$L$ &$N^c$&$H^D$&$g^c$    &$H^U$&$g$ &$S$ &$\Phi$&$\Phi^c$\\ \hline
$SU(3)_c$&$3$ &$3^*$    &$1$  &$3^*$    &$1$ &$1$  &$1$  &$3^*$    &$1$  &$3$ &$1$ &$1$   &$1$     \\ \hline
$SU(2)_W$&$2$ &$1$      &$1$  &$1$      &$2$ &$1$  &$2$  &$1$      &$2$  &$1$ &$1$ &$1$   &$1$     \\ \hline
$y=6Y$   &$1$ &$-4$     &$6$  &$2$      &$-3$&$0$  &$-3$ &$2$      &$3$  &$-2$&$0$ &$0$   &$0$     \\ \hline
$x$      &$1$ &$1$      &$1$  &$2$      &$2$ &$0$  &$-3$ &$-3$     &$-2$ &$-2$&$5$ &$0$   &$0$     \\ \hline
$z$      &$-1$&$-1$     &$-1$ &$2$      &$2$ &$-4$ &$-1$ &$-1$     &$2$  &$2$ &$-1$&$8$   &$-8$    \\ \hline
$R$      &$-$ &$-$      &$-$  &$-$      &$-$ &$-$  &$+$  &$+$      &$+$  &$+$ &$+$ &$+$   &$+$     \\ \hline
$q_\psi$ &$1$ &$1$      &$1$  &$1$      &$1$ &$1$  &$-2$ &$-2$     &$-2$ &$-2$&$4$ &$-2$  &$2$     \\
\hline
$q_\chi$ &$-1$&$-1$     &$-1$ &$3$      &$3$ &$-5$ &$-2$ &$-2$     &$2$  &$2$ &$0$ &$10$  &$-10$   \\
\hline
\end{tabular}
\end{center}
\caption{$G_2$ assignment of superfields.
Where the $x$, $y$ and $z$ are charges of $U(1)_X$, $U(1)_Y$ and $U(1)_Z$, and $Y$ is hypercharge.
The extra U(1) charges $x$ and $z$ are given by $x=\frac54 q_\psi+\frac14 q_\chi$ and
$z=-\frac14 q_\psi+\frac34 q_\chi$ , where $q_\psi=6 \sqrt{\frac25}Q_\psi$ and $q_\chi=2\sqrt{6}Q_\chi$ are charges of two U(1)s in
$SU(5)\times U(1)_\psi\times U(1)_\chi \subset SO(10)\times U(1)_\psi \subset E_6$.
}
\end{table}

We extend the gauge symmetry from $G_{SM}$ to $G_2=G_{SM}\times U(1)_X\times U(1)_Z\subset E_6$,
and add new superfields $S,g,g^c,N^c$ which are embedded in ${\bf 27}$ representation of $E_6$ with quark, lepton superfields $Q,U^c,D^c,L,E^c$ and Higgs superfields $H^U,H^D$.
In order to break $U(1)_Z$ gauge symmetry, we introduce
$G_{SM}$ singlet $\Phi$ and $\Phi^c$. The gauge representations of these superfields are given in Table 1.
After the gauge symmetry breaking, as the R-parity symmetry
\eqn{
R=\exp\L[\frac{i\pi}{20}(3x-8y+15z)\R]
}
remains unbroken, LSP is the candidate for dark matter.
The invariant superpotential under the gauge symmetry $G_2$ is given by
\eqn{
W&=&Y^UH^UQU^c+Y^DH^DQD^c+Y^EH^DLE^c \no \\
&+&\lambda SH^UH^D+kSgg^c \no \\
&+&Y^NH^ULN^c+Y^M\Phi N^cN^c \no \\
&+&M_\Phi\Phi^c\Phi+y_1QQg+y_2g^cU^cD^c+y_3gE^cU^c+y_4g^cLQ+y_5gD^cN^c,
}
where first line consists of trilinear terms in MSSM.
Second line generates effective $\mu$ term $\lambda\L<S\R>H^UH^D$ by
radiative symmetry breaking of $U(1)_X$.
Third line generates RHN mass term $Y^N\L<\Phi\R>N^cN^c$ by
radiative symmetry breaking of $U(1)_Z$ and gives small neutrino mass
by seesaw mechanism.
Fourth line consists of unwanted terms which cause the problems such that 
the mass term $M_\Phi\Phi\Phi^c$ prevents $\Phi,\Phi^c$ from 
developing vacuum expectation values (VEVs) and
the trilinear terms of exotic quarks destabilize proton.
Note that Higgs superfields are extended to three generations.
Generally, extra Higgs doublets cause the problem of flavor changing neutral currents (FCNCs).

\subsection{Flavor symmetry}

\begin{table}[htbp]
\begin{center}
\begin{tabular}{|c|c|c|c|c|c|c|c|c|c|}
\hline
        &$Q_1$    &$Q_2$    &$Q_3$     &$U^c_1$  &$U^c_2$  &$U^c_3$   &$D^c_1$  &$D^c_2$  &$D^c_3$ \\
      \hline
$S_4$   &${\bf 1}$&${\bf 1}$& ${\bf 1}$&${\bf 1}$&${\bf 1}$&${\bf 1}$ &${\bf 1}$&${\bf 1}$&${\bf 1}$\\
      \hline
$Z_2$   &$+$      &$+$      &$+$       &$+$      &$+$      &$+$       &$+$      &$+$      &$+$ \\
      \hline
      \hline
        &$E^c_1$  &$E^c_2$  &$E^c_3$   &$L_i$    &$L_3$    &$N^c_i$   &$N^c_3$  &$H^D_i$  &$H^D_3$  \\
      \hline
$S_4$   &${\bf 1}$&${\bf 1}$&${\bf 1'}$&${\bf 2}$&${\bf 1}$&${\bf 2}$ &${\bf 1}$&${\bf 2}$&${\bf 1}$ \\
      \hline      
$Z_2$   &$+$      &$-$      &$+$       &$-$      &$-$      &$+$       &$-$      &$-$      &$+$ \\
      \hline
      \hline   
        &$H^U_i$  &$H^U_3$  &$S_i$     &$S_3$    &$g_a$    &$g^c_a$   &$\Phi_i$ &$\Phi_3$ &$\Phi^c_a$\\
      \hline
$S_4$   &${\bf 2}$&${\bf 1}$&${\bf 2}$ &${\bf 1}$&${\bf 3}$&${\bf 3}$ &${\bf 2}$&${\bf 1}$&${\bf 3}$\\
      \hline
$Z_2$   &$-$      &$+$      &$-$       &$+$      &$+$      &$+$       &$-$      &$+$      &$+$ \\
      \hline
\end{tabular}
\end{center}
\caption{$S_4\times Z_2$ assignment of superfields
(Where the index $i$ of the $S_4$ doublets runs $i=1,2$,
and the index $a$ of the $S_4$ triplets runs $a=1,2,3$.)}
\end{table}

In ordre to explain maximal mixing angle $\theta_{23}$,
we introduce $S_4\times Z_2$ flavor symmetry. 
This symmetry solves the problems of superpotential defined in Eq.(2) at the same time.
If we assign $g,g^c,\Phi^c$ to $S_4$ -triplets and the other superfields to singlets or doublets,
then $M_\Phi,y_{1,\cdots,5}$ are eliminated.
As a result, superpotential is given by
\eqn{
W'&=&Y^UH^UQU^c+Y^DH^DQD^c+Y^EH^DLE^c \no \\
&+&\lambda SH^UH^D+kSgg^c \no \\
&+&Y^NH^ULN^c+Y^M\Phi N^cN^c \no \\
&+&\frac{a}{M_P}\Phi\Phi\Phi^c\Phi^c \no \\
&+&\frac{y'_1}{M^2_P}\Phi\Phi^cQQg
+\frac{y'_2}{M^2_P}\Phi\Phi^cg^cU^cD^c
+\frac{y'_3}{M^2_P}\Phi\Phi^cgE^cU^c\no \\
&+&\frac{y'_4}{M^2_P}\Phi\Phi^cg^cLQ
+\frac{y'_5}{M^2_P}\Phi\Phi^cgD^cN^c+\cdots,
}
where the dots $\cdots$ are higher order terms.
As the potential of $\Phi$ and $\Phi^c$
is lifted by non-renormalizable term $\Phi\Phi\Phi^c\Phi^c$,
$\Phi$ and $\Phi^c$ have very large VEVs along the D-flat direction of $\L<\Phi\R>=\L<\Phi^c\R>$,
where
\eqn{
V=\L<\Phi\R>=\L<\Phi^c\R>\sim \L(\frac{M_Pm_{SUSY}}{a}\R)^\frac12
\sim 10^{11}\mbox{GeV}\frac{1}{\sqrt{a}}\L(\frac{m_{SUSY}}{10\mbox{TeV}}\R)^\frac12.
}

From the constraints on the lifetimes of proton and exotic quarks (see appendix C), the condition
\eqn{
\frac{V^2}{M^2_P}\sim 10^{-12}
}
must be satisfied \cite{f-extra-u1}. From this condition, $V\sim 10^{12}\mbox{GeV}$ is required. This value is realized by potential minimum condition as Eq.(4),
when we take $a\sim 10^{-2}$. 
The prediction for RHN mass is given by
\eqn{
M_R\sim 10^{12}\mbox{GeV},
}
which gives an appropriate neutrino mass.
As the VEVs of $\Phi$ and $\Phi^c$ break not only $U(1)_Z$ but also
$S_4\times Z_2$, we call them flavons.

\subsection{Maki-Nakagawa-Sakata matrix $V_{MNS}$}

The maximal mixing angle $\theta_{23}$ of $V_{MNS}$ is realized by the assignments
that $H^U,H^D,L,N^c$ are ${\bf 2+1}$ of $S_4$ and $E^c$ is ${\bf 1+1+1'}$ \cite{kubo}.
In order to reduce the number of parameters, we assign $S$ and $\Phi$ to ${\bf 2+1}$.
If we assign $Q,U^c$ and $D^c$ to $S_4$-singlets, then quarks do not couple to
$S_4$ doublet Higgs and FCNC is suppressed. The flavor representations are given in Table 2. The leading order superpotential is given by
\eqn{
W_{S_4\times Z_2}&=&W_L+W_Q+W_H+W_g+W_\Phi , \\
W_L&=&Y^N_2\L[H^U_1(L_1N^c_2+L_2N^c_1)+H^U_2(L_1N^c_1-L_2N^c_2)\R] \no \\
&+&Y^N_3H^U_3L_3N^c_3+Y^N_4L_3(H^U_1N^c_1+H^U_2N^c_2) \no \\
&+&Y^E_1E^c_1(H^D_1L_1+H^D_2L_2)+Y^E_2E^c_2H^D_3L_3+Y^E_3E^c_3(H^D_1L_2-H^D_2L_1) \no \\
&+&\frac12 Y^M_1\Phi_3(N^c_1N^c_1+N^c_2N^c_2)+ \frac12 Y^M_3\Phi_3 N^c_3N^c_3 , \\
W_Q&=&Y^U_{ij}H^U_3Q_iU^c_j+Y^D_{ij}H^D_3Q_iD^c_j \quad(i,j=1,2,3) , \\
W_H&=&\lambda_1S_3(H^U_1H^D_1+H^U_2H^D_2)+\lambda_3S_3H^U_3H^D_3  \no \\
&+&\lambda_4H^U_3(S_1H^D_1+S_2H^D_2)+\lambda_5(S_1H^U_1+S_2H^U_2)H^D_3 , \\
W_g&=&kS_3(g_1g^c_1+g_2g^c_2+g_3g^c_3) , \\
W_\Phi&=&\frac{a_1}{2M_P}\Phi^2_3[(\Phi^c_1)^2+(\Phi^c_2)^2+(\Phi^c_3)^2] \no \\
&+&\frac{a_2}{2M_P}(\Phi^2_1+\Phi^2_2)[(\Phi^c_1)^2+(\Phi^c_2)^2+(\Phi^c_3)^2] \no \\
&+&\frac{a_3}{2M_P}\L\{2\sqrt{3}\Phi_1\Phi_2[(\Phi^c_2)^2-(\Phi^c_3)^2]
+(\Phi^2_1-\Phi^2_2)[(\Phi^c_2)^2+(\Phi^c_3)^2-2(\Phi^c_1)^2]\R\} .
}
We give the parameter set to realize the maximal mixing angle of MNS matrix.
We define non-negative VEVs as
\eqn{
&&\L<H^U_1\R>=\L<H^U_2\R>=\frac{1}{\sqrt{2}}v_u,\quad 
\L<H^U_3\R>=v'_u ,\quad 
\L<H^D_1\R>=\L<H^D_2\R>=\frac{1}{\sqrt{2}}v_d,\quad 
\L<H^D_3\R>=v'_d ,\no \\
&&\L<S_1\R>=\L<S_2\R>=\frac{1}{\sqrt{2}}v_s,\quad
\L<S_3\R>=v'_s ,\no \\
&&\L<\Phi_1\R>=v_4,\quad \L<\Phi_2\R>=\sqrt{v^2_1+v^2_2+v^2_3-v^2_4},\quad
\L<\Phi_3\R>=V , \no \\
&&\L<\Phi^c_1\R>=\sqrt{\frac{V^2}{3}+v^2_1},\quad 
\L<\Phi^c_2\R>=\sqrt{\frac{V^2}{3}+v^2_2},\quad 
\L<\Phi^c_3\R>=\sqrt{\frac{V^2}{3}+v^2_3} \quad (v_{1,2,3,4}\ll V).
}
In $W_L$, without loss of generality, we can define $Y^N_{2,4},Y^E_{1,2,3},Y^M_{1,3}$ to be real and define the phase of $Y^N_3$ as $Y^N_3=|Y^N_3|e^{i\delta}$.
We define mass parameters as
\eqn{
\begin{tabular}{llll}
$M_1=Y^M_1V$,        & $M_3=Y^M_3V$,          &                     &   \\
$m^\nu_2=Y^N_2v_u$,  & $m^\nu_3=|Y^N_3|v'_u$, & $m^\nu_4=Y^N_4v_u$, &   \\
$m^l_1=Y^E_1v_d$,    & $m^l_2=Y^E_2v'_d$,     & $m^l_3=Y^E_3v_d$.   &
\end{tabular}
}
Using these parameters, the mass matrices of charged leptons and neutrinos are given by
\eqn{
\begin{tabular}{ll}
$M_l=\frac{1}{\sqrt{2}}\Mat3{m^l_1}{0}{-m^l_3}{m^l_1}{0}{m^l_3}{0}{\sqrt{2}m^l_2}{0}$, &
$M_D=\frac{1}{\sqrt{2}}\Mat3{m^\nu_2}{m^\nu_2}{0}{m^\nu_2}{-m^\nu_2}{0}
{m^\nu_4}{m^\nu_4}{\sqrt{2}e^{i\delta}m^\nu_3}$,  \\
$M_R=\Mat3{M_1}{0}{0}{0}{M_1}{0}{0}{0}{M_3}$. &
\end{tabular}
}
The following neutrino mass matrix is generated through the seesaw mechanism
\eqn{
M_\nu&=&M_DM^{-1}_RM^t_D=\Mat3{\rho^2_2}{0}{\rho_2\rho_4}
{0}{\rho^2_2}{0}
{\rho_2\rho_4}{0}{\rho^2_4+e^{2i\delta}\rho^2_3},
}
where
\eqn{
\rho_2=\frac{m^\nu_2}{\sqrt{M_1}},\quad \rho_4=\frac{m^\nu_4}{\sqrt{M_1}},\quad \rho_3=\frac{m^\nu_3}{\sqrt{M_3}}.
}
The mass eigenvalues and diagonalization matrix of charged leptons are given by
\eqn{
V^\dagger_l M^*_l M^t_l V_l&=&diag(m^2_e,m^2_\mu, m^2_\tau)=((m^l_2)^2,(m^l_3)^2,(m^l_1)^2), \\
V_l&=&\frac{1}{\sqrt{2}}\Mat3{0}{-1}{1}
{0}{1}{1}
{-\sqrt{2}}{0}{0},
}
and those of neutrinos are given by
\eqn{
V^t_\nu M_\nu V_\nu&=&diag(e^{i(\phi_1-\phi)}m_{\nu_1},e^{i(\phi_2+\phi)}m_{\nu_2},m_{\nu_3}), \\
V_\nu&=&
\Mat3{-\sin\theta_\nu}{e^{i\phi}\cos\theta_\nu}{0}
{0}{0}{1}
{e^{-i\phi}\cos\theta_\nu}{\sin\theta_\nu}{0}.
}
From Eq.(19) and Eq.(21), we obtain the MNS matrix as follows
\eqn{
V_{MNS}&=&V^\dagger_lV_\nu P_\nu
=\frac{1}{\sqrt{2}}\Mat3{-\sqrt{2}e^{-i\phi}\cos\theta_\nu}{-\sqrt{2}\sin\theta_\nu}{0}
{\sin\theta_\nu}{-e^{i\phi}\cos\theta_\nu}{1}
{-\sin\theta_\nu}{e^{i\phi}\cos\theta_\nu}{1}P_\nu,
}
where
\eqn{
P_\nu=\mbox{diag}(e^{-i(\phi_1-\phi)/2},e^{-i(\phi_2+\phi)/2},1).
}
Here it is worth mentioning that the lower bound of $(0.04<)\theta_{13}$ was shown by  the recent experiment reported by T2K \cite{Abe:2011sj} at 90 $\%$ C.L., which could give a severe test to our model near future.

From the experimental bound \cite{PDG2008}, we impose the condition
\eqn{
\tan\theta_\nu=\frac{1}{\sqrt{2}},\quad
m^2_{\nu_2}-m^2_{\nu_1}=8.0\times 10^{-5}(\mbox{eV}^2),\quad
m^2_{\nu_2}-m^2_{\nu_3}=2.5\times 10^{-3}(\mbox{eV}^2),
}
on the parameters, then the phase $\phi$ is given by
\eqn{
r\cos\phi =0.361,\quad r=\frac{\rho_2}{\rho_4}.
}
Fixing the VEVs as
\eqn{
v_u=10,\quad v'_u=155.3,\quad v_d=2.0,\quad v'_d=77.8\quad (\mbox{GeV}),
\label{vev}
}
and the charged lepton masses as \cite{s4e6}\cite{f-mass}
\eqn{
m^l_1=1.75\mbox{GeV},\quad  m^l_2=487\mbox{keV}, \quad m^l_3=103\mbox{MeV},
}
Yukawa coupling constants are given by
\eqn{
Y^E_1=0.875,\quad Y^E_3=5.15\times 10^{-2},\quad Y^E_2=6.25\times 10^{-6}.
}
For the RHN mass parameters, we assume 
\eqn{
V=10^{12}\mbox{GeV},\quad Y^M_1=Y^M_3=1.
}
In order to investigate model dependence, we give two sample parameter sets A and B 
which are defined as follows
\eqn{
\mbox{A}&:&r=0.361 \no \\
&&\phi=\phi_1=\phi_2=0.0^\circ,\quad \delta=90.0^\circ ,\no \\
&&\rho^2_2=1.80\times 10^{-2}\mbox{eV},\quad
\rho^2_3=14.47\times 10^{-2}\mbox{eV},\quad \rho^2_4=13.78\times 10^{-2}\mbox{eV} , \no \\
&&m_{\nu_1}=5.24\times 10^{-2}\mbox{eV},\quad 
m_{\nu_2}=5.31\times 10^{-2}\mbox{eV},\quad m_{\nu_3}=1.80\times 10^{-2}\mbox{eV} , \no \\
&&m^\nu_2=4.24\mbox{GeV},\quad m^\nu_3=12.0 \mbox{GeV},\quad m^\nu_4=11.7\mbox{GeV}, \no  \\
&&Y^N_2=0.424,\quad Y^N_3=0.077,\quad Y^N_4=1.17, \\
\mbox{B}&:&r=1.000 \no \\
&&\phi=68.84^\circ, \quad \phi_1=41.55^\circ ,\quad \phi_2=41.15^\circ ,\quad \delta=89.805^\circ ,\no \\
&&\rho^2_2=5.03\times 10^{-2} \mbox{eV},\quad \rho^2_3=10.04\times 10^{-2} \mbox{eV}, 
\quad \rho^2_4=5.03\times 10^{-2}\mbox{eV}, \no \\
&&m_{\nu_1}=7.08\times 10^{-2} \mbox{eV},\quad m_{\nu_2}=7.14\times 10^{-2} \mbox{eV},\quad 
m_{\nu_3}=5.03\times 10^{-2} \mbox{eV} \no \\
&&m^\nu_2=7.09\mbox{GeV},\quad m^\nu_3=10.02 \mbox{GeV},\quad m^\nu_4=7.09\mbox{GeV} , \no \\
&&Y^N_2=0.709,\quad Y^N_3=0.065,\quad Y^N_4=0.709.
}
Generally, multi-Higgs model causes FCNC problems, however our assignments do not
cause such problems. In the lepton sector, the interactions between
charged leptons and Higgs bosons are given by
\eqn{
{\cal L}_l
&=&Y^E_1\tau^c\L[l_\mu\L(\frac{H^D_2-H^D_1}{\sqrt{2}}\R)
+l_\tau\L(\frac{H^D_1+H^D_2}{\sqrt{2}}\R)\R]
-Y^E_2H^D_3e^cl_e \no \\
&+&Y^E_3\mu^c\L[l_\mu\L(\frac{H^D_1+H^D_2}{\sqrt{2}}\R)
+l_\tau\L(\frac{H^D_1-H^D_2}{\sqrt{2}}\R)\R],
}
which do not contribute to $\tau\to e+\gamma,\mu\to e+\gamma$ processes.
Because ${\cal L}_l$ has accidental $S_2$ symmetry such as
\eqn{
(H^D_1,H^D_2)\to (H^D_2,H^D_1),\quad (l_\mu,\mu^c)\to (-l_\mu,-\mu^c),
}
$\tau\to\mu+\gamma$ process is also not induced.
Note that this $S_2$ symmetry is not the symmetry of whole theory, 
as the symmetry is violated in neutrino Yukawa couplings and flavon superpotential $W_\Phi$.

In the quark sector, as the quarks couple only to $H^U_3,H^D_3$,
Higgs mediated FCNCs are not induced.  In the basis that quark mass matrices are diagonal, the superpotential is written as
\eqn{
W_Q&=&Y_tH^U_3(Q_3)'(U^c_3)'+Y_cH^U_3(Q_2)'(U^c_2)'+Y_uH^U_3(Q_1)'(U^c_1)' \no \\
&+&Y_bH^D_3(Q_3)'(D^c_3)'+Y_sH^D_3(Q_2)'(D^c_2)'+Y_dH^D_3(Q_1)'(D^c_1)' .
}
From here, we fix top, bottom and charm masses as \cite{s4e6}\cite{f-mass}
\eqn{
Y_tv'_u=172.5,\quad Y_bv'_d=2.89,\quad Y_cv'_u=0.624\quad (\mbox{GeV}),
}
and assume exotic quark mass as
\eqn{
kv'_s=2000\quad (\mbox{GeV}),
}
in order to forbid the decay of lightest flavon into exotic quark pair.
Then we fix the values of Yukawa coupling constants as
\eqn{
Y_t=1.12,\quad Y_b=0.0371,\quad Y_c=0.00405,\quad k=1.0,\quad v'_s=2000\mbox{GeV},\quad
v_s=200\mbox{GeV}.
}

\subsection{Higgs sector}

Higgs potential is given as follows,
\eqn{
V&=&V_F+V_D+V_A+V_{m^2} ,   \\
V_F&=&\L|\lambda_1S_3H^D_1+\lambda_5S_1H^D_3\R|^2
+\L|\lambda_1S_3H^D_2+\lambda_5S_2H^D_3\R|^2 \no \\
&+&\L|\lambda_3S_3H^D_3+\lambda_4(S_1H^D_1+S_2H^D_2)\R|^2 \no \\
&+&\L|\lambda_1S_3H^U_1+\lambda_4S_1H^U_3\R|^2
+\L|\lambda_1S_3H^U_2+\lambda_4S_2H^U_3\R|^2 \no \\
&+&\L|\lambda_3S_3H^U_3+\lambda_5(S_1H^U_1+S_2H^U_2)\R|^2 \no \\
&+&\L|\lambda_4H^U_3H^D_1+\lambda_5H^U_1H^D_3\R|^2
+\L|\lambda_4H^U_3H^D_2+\lambda_5H^U_2H^D_3\R|^2 \no \\
&+&\L|\lambda_1(H^U_1H^D_1+H^U_2H^D_2)+\lambda_3H^U_3H^D_3\R|^2 , \\
V_D&=&\frac12g^2_Y\L(\frac12|H^U_a|^2-\frac12|H^D_a|^2\R)^2
+\frac12g^2_2\sum_A\L((H^U_a)^\dagger T^AH^U_a+(H^D_a)^\dagger T^A H^D_a\R)^2 \no \\
&+&\frac12g^2_x\L(-2|H^U_a|^2-3|H^D_a|^2+5|S_a|^2\R)^2 , \\
V_A&=&-\lambda_1A_1S_3(H^U_1H^D_1+H^U_2H^D_2)-\lambda_3A_3S_3H^U_3H^D_3\no \\
&-&\lambda_4A_4H^U_3(S_1H^D_1+S_2H^D_2)-\lambda_5A_5(S_1H^U_1+S_2H^U_2)H^D_3+h.c.  , \\
V_{m^2}&=&-m^2_{U3}|H^U_3|^2+m^2_U(|H^U_1|^2+|H^U_2|^2)
+m^2_{D3}|H^D_3|^2+m^2_D(|H^D_1|^2+|H^D_2|^2) \no \\
&-&m^2_{S3}|S_3|^2+m^2_S(|S_1|^2+|S_2|^2) \no \\
&-&\L[m^2_{BU}(H^U_3)^\dagger(H^U_1+H^U_2)
+m^2_{BD}(H^D_3)^\dagger(H^D_1+H^D_2)+m^2_{BS}(S_3)^\dagger(S_1+S_2)+h.c.\R], 
}
where we can define $\lambda_{1,3,4,5}$ to be real without loss of generality,
and we assume all the soft SUSY breaking parameters are real to avoid complex VEVs.

he soft $S_4\times Z_2$ breaking terms; $m^2_{BU}, m^2_{BD},m^2_{BS}$, violate 
accidental $O(2)$ symmetry of Higgs potential and fix the VEV directions (Eq.(13))
to realize $\theta_{23}=45^\circ$. This potential has $S_2$ symmetry such as
\eqn{
H^U_1\leftrightarrow H^U_2,\quad
H^D_1\leftrightarrow H^D_2,\quad
S_1\leftrightarrow S_2.
}
Minimizing this potential, we get mass matrices of Higgs bosons.
The results are given in appendix A.
In the same manner, we add soft $S_4\times Z_2$ breaking terms in flavon sector
to avoid domain wall problem \cite{domainwall}\footnote{We would like to thank Refree for the suggestion.}.

\section{Dark Matter}
Here we show that our model is consistent with cosmic-ray observation of PAMELA.
Decaying dark matter scenarios with Non-Abelian discrete flavor symmetries have been done by Ref. \cite{kaji}. 
\subsection{LF decay width}

We assume that the candidate for decaying dark matter is the lightest flavon (LF).
In the six flavon superfields; $\Phi_a,\Phi^c_a(a=1,2,3)$, only one linear combination
is super-heavy and the other five superfields have TeV scale masses.
As LF cannot decay into other flavons, it has very long lifetime.
Due to the non-renormalizable interactions with light particles, 
LF becomes unstable dark matter. Among the interactions,
the source term of RHN mass
\eqn{
W_{\mbox{eff}}=\frac{(Y^NH^UL)^2}{2Y^M(V+\Phi_3)}
\sim \frac{(Y^NH^UL)^2}{2M_R}\L(1-\frac{\Phi_3}{V}\R)
=\frac12 m_\nu\L(\frac{H^UL}{v}\R)^2\L(1-\frac{\Phi_3}{V}\R)
}
is the unique interaction to emit leptons without emitting quarks
\cite{nupamela}, where $v$ is VEV of $H^U$.
We estimate the positron flux using this interaction.
Due to the factor $1/v$, the Higgs which develops the smallest VEV 
gives the largest contribution to LF decay.
Therefore we can neglect the contribution from $H^U_3$, because $v_u\ll
v'_u$ as one can see from Eq.(\ref{vev}).
This effect is impotant to suppress weak boson emission.
Due to the enhancement factor $m_{LF}/v_u$, LF decay width is dominated by
4-body decay as follows
\eqn{
\Gamma(LF\to \nu+\nu)&\ll& \Gamma(LF\to \nu+l+H^+)
\sim \Gamma(LF\to \nu+l+W^+) \no \\
&\ll& \Gamma(LF\to l+l+H^++H^+).
}
From the spectrum of positron flux observed by PAMELA, we assume
\eqn{
m_{LF}=4\mbox{TeV}.
}
If we assume all sfermions which couple to LF are heavier than 4TeV,
the other interactions do not contribute to LF decay.
The interactions which contribute to LF decay is given as follows
\eqn{
{\cal L}_{2\nu}&=&\frac12C_{LF}\phi_{LF}
\L\{
[(H^U_1)^0(H^U_1)^0+(H^U_2)^0(H^U_2)^0](\nu_e\nu_e+r^2\nu_\mu\nu_\mu+r^2\nu_\tau\nu_\tau)
\R. \no \\
&+&\L. 2\sqrt{2}r(\nu_\mu-\nu_\tau)(H^U_1)^0(H^U_2)^0\nu_e
-\sqrt{2}r(\nu_\mu+\nu_\tau)[(H^U_1)^0(H^U_1)^0-(H^U_2)^0(H^U_2)^0]\nu_e \R\} , \\
{\cal L}_{l\nu}&=&-\frac12C_{LF}\phi_{LF}\L\{
2[(H^U_1)^0(H^U_1)^++(H^U_2)^0(H^U_2)^+](e\nu_e+r^2\nu_\mu\mu+r^2\nu_\tau\tau) \R. \no \\
&+&2r^2[(H^U_1)^0(H^U_2)^+-(H^U_2)^0(H^U_1)^+](\nu_\mu\tau-\nu_\tau\mu) \no \\
&+&\sqrt{2}r[(H^U_1)^0(H^U_2)^++(H^U_2)^0(H^U_1)^+]
[(\nu_\mu-\nu_\tau)e+\nu_e(\mu-\tau)] \no \\
&-&\L. \sqrt{2}r[(H^U_1)^0(H^U_1)^+-(H^U_2)^0(H^U_2)^+]
[(\nu_\mu+\nu_\tau)e+\nu_e(\mu+\tau)]
\R\} , \\
{\cal L}_{2l}&=&\frac12C_{LF}\phi_{LF}
\L\{
[(H^U_1)^+(H^U_1)^++(H^U_2)^+(H^U_2)^+](ee+r^2\mu\mu+r^2\tau\tau)
\R. \no \\
&+&\L. 2\sqrt{2}r(\mu-\tau)(H^U_1)^+(H^U_2)^+e
-\sqrt{2}r(\mu+\tau)[(H^U_1)^+(H^U_1)^+-(H^U_2)^+(H^U_2)^+]e \R\} , \\
C_{LF}&=&\frac{\epsilon\rho^2_4}{\sqrt{2}Vv^2_u},
}
where $\epsilon$ is the flavon mixing parameter which is defined by
\eqn{
\Phi_3=\frac{\epsilon\phi_{LF}}{\sqrt{2}},
}
where $\phi_{LF}$ is LF field.

\begin{figure}[ht]
\unitlength=1mm \hspace{5cm}
\begin{picture}(70,50)
\includegraphics[height=5cm,width=7cm]{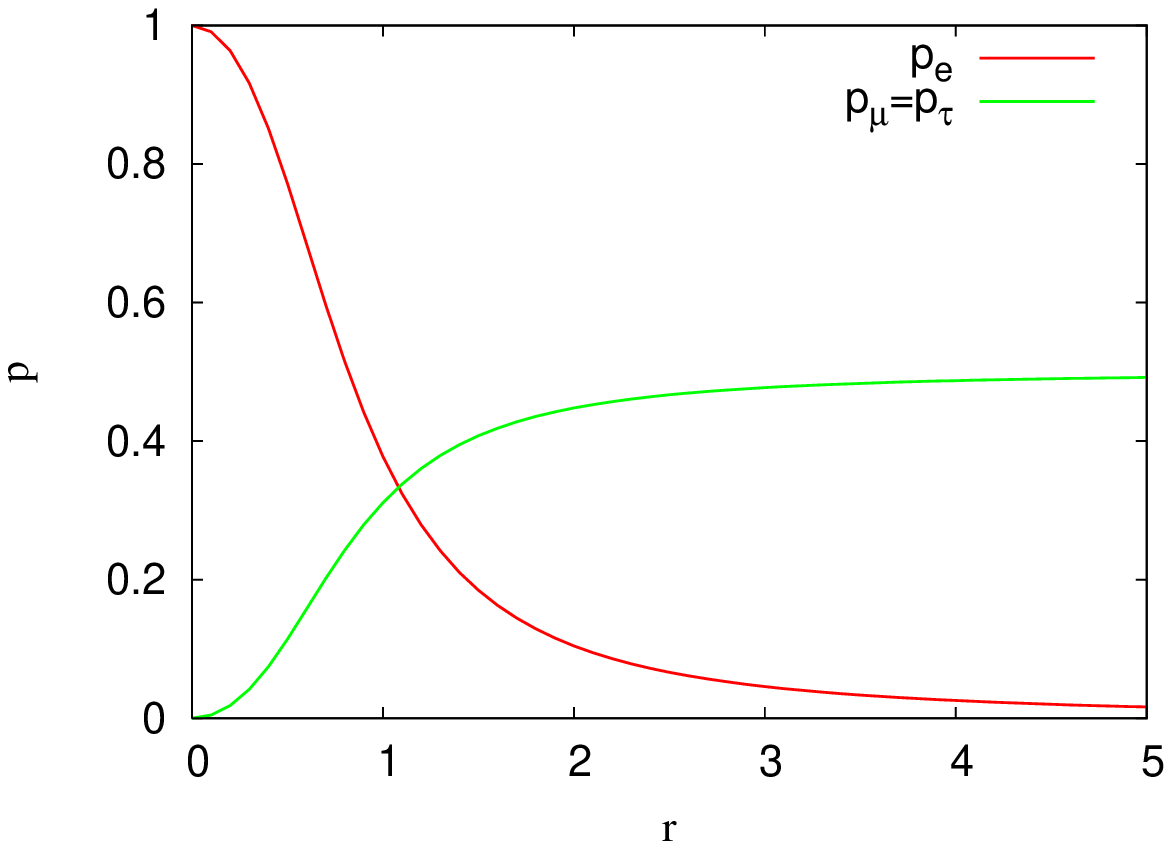}
\end{picture}
\caption{}
\label{r-p}
\end{figure}

Using Eq.(47)-(49), the LF decay widths are given as follows
\eqn{
\Gamma_{2\nu}&=&\Gamma_{2\bar{\nu}}=(6+10r^2+12r^4)\Gamma_0 ,\\
\Gamma_{l\nu}&=&\Gamma_{\bar{l}\bar{\nu}}=\Gamma_e+\Gamma_\mu+\Gamma_\tau
=(2+8r^2+8r^4)\Gamma_0 ,\no \\
&&\Gamma_e=\Gamma_{\bar{e}}=(2+4r^2)\Gamma_0 ,\quad
\Gamma_\mu=\Gamma_{\bar{\mu}}=\Gamma_\tau=\Gamma_{\bar{\tau}}
=(2r^2+4r^4)\Gamma_0 ,\no \\
\Gamma_{2l}&=&\Gamma_{2\bar{l}}
=\Gamma_{2e}+\Gamma_{2\mu}+\Gamma_{2\tau}+\Gamma_{e\mu}+\Gamma_{e\tau}
=(8+12r^2+16r^4)\Gamma_0 ,\\
&&\Gamma_{2e}=8\Gamma_0 ,\quad
\Gamma_{2\mu}=\Gamma_{2\tau}=8r^4\Gamma_0  ,\quad
\Gamma_{e\mu}=\Gamma_{e\tau}=6r^2\Gamma_0 ,\\
\Gamma_{\mbox{lepton}}&=&\Gamma_{l\nu}+\Gamma_{2l}=(10+20r^2+24r^4)\Gamma_0
=\Gamma_{\mbox{anti-lepton}} ,\no \\
\Gamma_{\mbox{total}}&=&2(\Gamma_{2\nu}+\Gamma_{l\nu}+\Gamma_{2l})
=2(16+30r^2+36r^4)\Gamma_0 ,\\
\Gamma_0&=&\frac{m_{LF}}{16\pi}\L(\frac{m^2_{LF}\epsilon\rho^2_4}{32\pi^2v^2_uV}\R)^2(0.111),
}
where we classify the final states only by charged lepton flavor $e,\mu,\tau$.
The rates of lepton flavor emitted by LF decay are given by
\eqn{
p_e=\frac{9+8r^2}{9+16r^2+20r^4},\quad
p_\mu=p_\tau=\frac{4r^2+10r^4}{9+16r^2+20r^4}.
\label{p_ell}
}
Anti-lepton flux depends on Majorana phase $\phi$ through Eq.(25), such as
e-dominant for $r<1$ and $(\mu,\tau)$-dominant for $r>1$ (see Fig. \ref{r-p}).
For each parameter set, LF lifetime is estimated as follows
\eqn{
\mbox{A}&:&
\Gamma^{-1}_{\mbox{total}}=3.72\times 10^{11}\epsilon^{-2}\mbox{sec},\quad
\Gamma^{-1}_{\mbox{anti-lepton}}=1.17\times 10^{12}\epsilon^{-2}\mbox{sec} , \\
\mbox{B}&:& 
\Gamma^{-1}_{\mbox{total}}=7.01\times 10^{11}\epsilon^{-2}\mbox{sec},\quad 
\Gamma^{-1}_{\mbox{anti-lepton}}=2.13\times 10^{12}\epsilon^{-2}\mbox{sec}.
}
Hereafter we assume $\epsilon< 10^{-3}$ to avoid extinction of LF. 

\subsection{Relic Abundance of LF}

At early stage of the universe, flavon multiplets are produced through 
$U(1)_Z$ gauge interaction \cite{u1-pamela}.
Since we assume that reheating temperature is low enough to avoid
gravitino over-production as $T_{RH}<10^7\mbox{GeV}$ \cite{gravitino},
this interaction is never thermal equiliburium.
Therefore we assume non-thermal production of flavons and
boundary condition $n_{LF}(T_{RH})=0$.

For the chiral multiplets $(\psi_L,\Psi)$,
$U(1)_Z$ gauge interaction is given by
\eqn{
{\cal L}_{U(1)_Z}=ig_zA^\mu\sum_iz_i\L[\bar{\psi}_{i,L}\gamma_\mu\psi_{i,L}
+\Psi_i\pl_\mu\Psi^\dagger_i-\Psi^\dagger_i\pl_\mu\Psi_i\R],
}
from which we calculate  production cross sections of flavon multiplets $(\phi,\Phi)$.
From Eq.(13), the $U(1)_Z$ gauge boson mass is nearly equal to $16g_z V$.

As all produced flavon multiplets decay into LF finally,
LF number density is given by
\eqn{
n_{LF}=5N_{LF}(n_\phi+n_\Phi),
}
where $5$ is the number of light flavon superfields,
$n_\Phi,n_\phi$ are number density of one flavon multiplet
and $N_{LF}\sim O(1)$ is LF production rate
which means how many LFs are produced per one degree of freedom
of flavon multiplets. The Boltzmann equation for
$n_{LF}$ is given by
\eqn{
\dot{n}_{LF}+3Hn_{LF}&=&2480N_{LF}CT^8, \\
C&=&\frac{21}{(2\pi)^5}\L(\frac{1}{32V^2}\R)^2,
}
from which we get
\eqn{
\Omega_{LF}h^2&=&\frac{m_{LF}s_0h^2}{\rho_c}\L[\frac{15\times 2480\times 21 m_PN_{LF}}
{2\pi^2(341.25)\times 30.67(2\pi)^5}
\L(\frac{1}{32V^2}\R)^2T^3_{RH}\R] \no \\
&=&5.06\times 10^{-9}N_{LF}\L(\frac{T_{RH}}{10^5GeV}\R)^3,
}
where \cite{PDG2008}
\eqn{
H&=&1.66\sqrt{g_*}\frac{T^2}{m_P}, \no \\
g_*&=&341.25 ,\no \\
m_P&=&1.22\times 10^{19}\mbox{GeV}, \no \\
s_0&=&2890/\mbox{cm}^3 ,\no \\
\rho_c&=&1.05\times 10^4h^2\mbox{eV}/\mbox{cm}^3.
}
For $T_{RH}<10^7\mbox{GeV}$, LF does not dominate dark matter
($\Omega_{LF}h^2\ll \Omega_{DM}h^2=0.11$), thus other dark matter should
be considered as we will discuss later. Such multi-component dark matter is discussed in \cite{Cao:2007fy}.
Although the number density is very low,
the short lifetime of LF enables us to explain cosmic-ray observation.
The effective lifetime of LF is defined as
\eqn{
\tau_{\mbox{eff}}\equiv\Gamma^{-1}_{\mbox{anti-lepton}}\L(\frac{\Omega_{DM}}{\Omega_{LF}}\R),
}
and the following values are obtained for each parameter set
\eqn{
\mbox{A}&:&\tau_{\mbox{eff}}=6.0\times 10^{25}\mbox{sec} ,\quad
\epsilon^2N_{LF}\L(\frac{T_{RH}}{10^5\mbox{GeV}}\R)^3=4.2\times 10^{-7},\\
\mbox{B}&:&\tau_{\mbox{eff}}=7.0\times 10^{25}\mbox{sec},\quad
\epsilon^2N_{LF}\L(\frac{T_{RH}}{10^5\mbox{GeV}}\R)^3=6.6\times 10^{-7}.
}
Eq.(67) and Eq.(68) are satisfied for example, if we put
$N_{LF}\sim 1,T_{RH}\sim 10^5\mbox{GeV},\epsilon\sim 10^{-3}$.
\\ \\

\begin{figure}[ht]
\begin{center}
\unitlength=1mm \hspace{1cm}
\includegraphics[height=5.3cm]{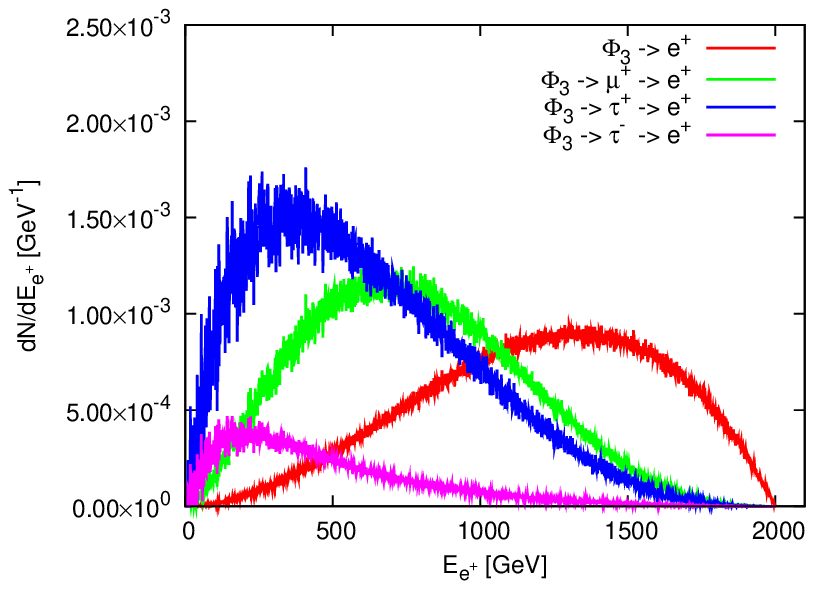}
\includegraphics[height=5.3cm]{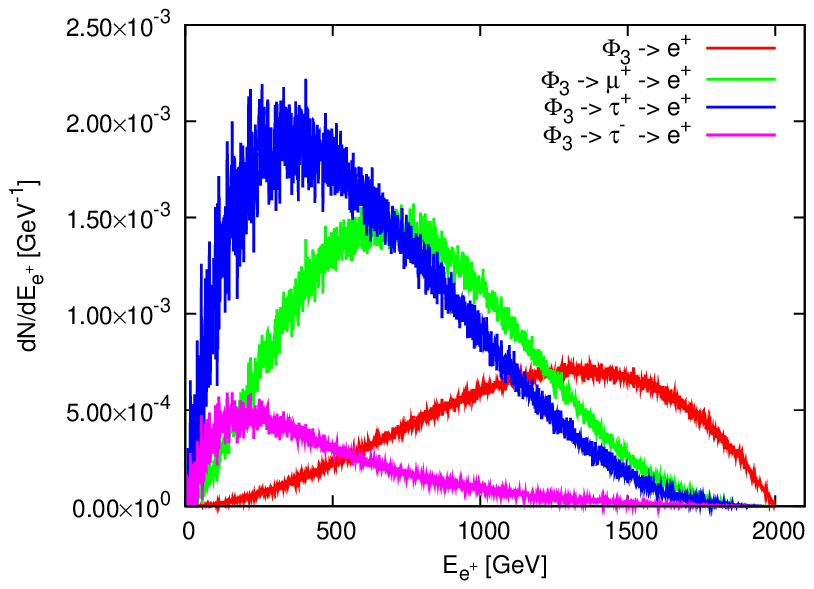}
\caption{The fragmentation function calculated by \texttt{pythia} for
 parameter set A (left) and B (right).}
\label{fragmentation}
\end{center}
\end{figure}

\begin{figure}[ht]
\begin{center}
\unitlength=1mm \hspace{1cm}
\includegraphics[height=5.3cm]{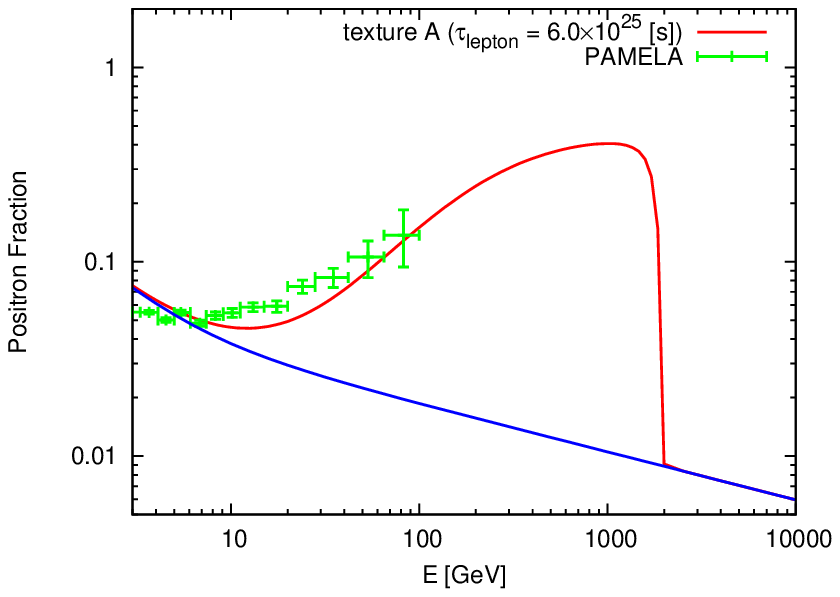}
\includegraphics[height=5.3cm]{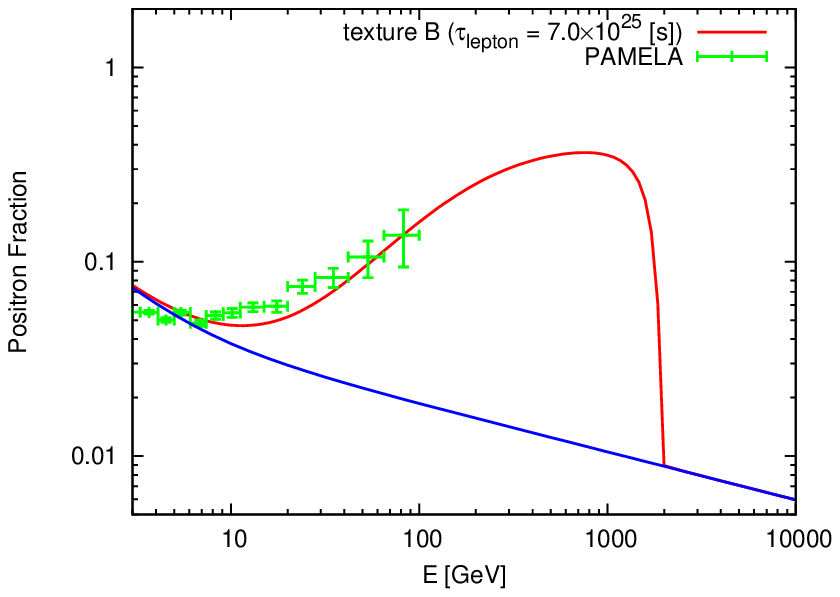}
\caption{ The positron flux calculated for parameter set A (left) and B (right).}
\label{pam}
\end{center}
\end{figure}

The positron flux from the decay of LF is calculated as 
\begin{eqnarray}
\Phi(E_{e^+})=\frac{v_{e^+}}{4\pi}\frac{1}{m_{LF}\tau_{\mathrm{eff}}}\int dE'
G_{e^+}(E_{e^+},E')\sum_{\ell=e^+,\mu^+,\tau^\pm}p_{\ell}\frac{dN_{\ell\: e^+}}{dE'},
\end{eqnarray}
where $v_{e^+}$ is the velocity of the positron, $G_{e^+}$ is the
Green's fuction which is expressed in \cite{Ibarra:2008jk}, $p_{\ell}$
is expressed in Eq.(\ref{p_ell}) and $dN_{\ell\: e^+}/dE'$ is the
fragmentation function produced from the decay of $\ell$ to $e^+$. The
fragmentation function is calculated by using the event generator
$\texttt{pythia}$ \cite{Sjostrand:2006za} and the result is shown in
Fig.\ref{fragmentation}. 
We can evalutate the positron flux from the decay of LF by using the
fragmentation function. The results for each parameter set A and B are 
shown in Fig.\ref{pam}. 

From the gamma-ray observations \cite{gamma}, the constraint for
$\tau$-flux is given by
\eqn{
(\tau_\tau)_{\mbox{eff}}
=\frac{\Gamma_{\mbox{lepton}}}{\Gamma_\tau}\tau_{\mbox{eff}}
=\frac{10+20r^2+24r^4}{8r^2+20r^4}\tau_{\mbox{eff}}
\geq 2.1\times 10^{26}\mbox{sec},
}
which is estimated for each parameter set as follows
\eqn{
\mbox{A}&:&(\tau_\tau)_{\mbox{eff}}=5.6\times 10^{26}\mbox{sec}, \no \\
\mbox{B}&:&(\tau_\tau)_{\mbox{eff}}=1.4\times 10^{26}\mbox{sec}.
}
As the parameter set B is severe to satisfy Eq.(70),
small $r$ model is favored.
This is the new information about neurtino sector extarcted by cosmic-ray observations.

\subsection{Higgs decay width}

No excess of anti-proton flux in cosmic-ray constrains
the species of the particles emitted by LF decay \cite{Adriani:2008zq}.
As the weak boson $Z,W^\pm$ and the chargino decay mainly into quarks,
LF should not decay into these particles so much.
The weak boson emission is suppressed by factor $(v_u/m_{LF})^2$ and
the chargino emission channel is kinematically closed for heavy sfermion scenario.
In order to forbid the weak boson and the chargino emission from Higgs boson decay,
we assume light Higgs scenario.

In the Higgs potential Eqs.(38)-(42) and
mass terms of the neutralinos and the charginos
\eqn{
{\cal L}&\supset&-i\sqrt{2}(H^U_a)^\dagger[g_2\lambda^A_2T^A_2+3g_y\lambda_Y-2g_x\lambda_X]h^U_a \no \\
&-&i\sqrt{2}(H^D_a)^\dagger[g_2\lambda^A_2T^A_2-3g_y\lambda_Y-3g_x\lambda_X]h^D_a
-i\sqrt{2}(S_a)^\dagger[5g_x\lambda_X]s_a\no \\
&-&\frac12M_2\lambda^A_2\lambda^A_2-\frac12M_Y\lambda_Y\lambda_Y
-\frac12M_X\lambda_X\lambda_X+(W_H)_{\theta^2}+h.c. \quad (A=1,2,3),
}
we assume the parameters as follows
\eqn{
&&g_2=0.652,\quad g_y=\frac16 g_Y\quad  (g_Y=0.357), \quad g_x=\frac{1}{2\sqrt{6}}g_Y=0.073, \no \\
&&\lambda_1=0.065,\quad \lambda_3=0.4,\quad \lambda_4=0.398,\quad \lambda_5=0.75, \no \\
&&M_X=M_2=200,\quad M_Y=180 \quad (\mbox{GeV}) ,\no \\
&&A_1=0.0,\quad A_3=A_4=A_5=1.0\quad (\mbox{TeV}) ,\no \\
&&m^2_{BU}=0.0408,\quad m^2_{BD}=m^2_{BS}=0.02\quad (\mbox{TeV}^2).
}
Mass matrices of neutralinos and charginos are given in appendix A and the
values of mass eigenvalues and mixing matrices are given in appendix B.

We consider only the mass eigenstates which dominate $H^U_{1,2}$ such as
\eqn{
&&\phi'_1(91.50),\quad \phi'_4(121.96),\quad \phi'_5(152.48),\quad \rho'_1(112.12)
,\quad \rho'_6(145.53) ,\no \\
&&(H^\pm_1)'(90.70),\quad (H^\pm_3)'(130.02) \quad (\mbox{GeV}),
}
where $\phi'_1,\rho'_1,(H^\pm_1)'$ are $S_2$-odd and the others are even.
As $S_2$ forbids interaction $\phi'_1ZZ$ and $\phi'_1$ is not emitted 
through $Z^* \to Z+\phi'_1$,
LEP bound $m_H\geq 114.4\mbox{GeV}$ is not imposed on $\phi'_1$.
As the masses of these Higgs bosons are well degenerated,
they do not emit weak bosons or charginos.

The neutralinos into which
these Higgs bosons can decay are two singlino dominant neutralinos
\eqn{
\eta'_1(41.92),\quad \eta'_4(44.55)\quad (\mbox{GeV}),
}
where $\eta'_1$ is $S_2$-odd and LSP. 
$S_2$-even neutralino $\eta'_4$ can decay into $\eta'_1$ through
$\eta'_4\to \eta'_1+\mu+\bar{\tau}$ without emitting quarks.
As $S_2$-odd Higgs boson can not decay into quarks,
we consider only the decay of $S_2$-even Higgs bosons.

The decay widths of $\phi'_4,\phi'_5, \rho'_6$ due to the Yukawa interactions
\eqn{
{\cal L}&\supset&Y_c(H^U_3)^0cc^c+Y_b(H^D_3)^0bb^c
+Y^E_1[(H^D_1)^0e^c_1e_1+(H^D_2)^0e^c_1e_2]+h.c. ,
}
are given as follows
\eqn{
\Gamma(\phi'_4\to c+\bar{c})&=&3.63\times 10^{-6}\Gamma_{2d} ,\\
\Gamma(\phi'_4\to b+\bar{b})&=&1.54\times 10^{-4}\Gamma_{2d} ,\\
\Gamma(\phi'_4\to\tau+\bar{\tau})&=&1.10\times 10^{-6}\Gamma_{2d} ,\\
\Gamma(\phi'_5\to c+\bar{c})&=&1.61\times 10^{-5}\Gamma_{2d} ,\\
\Gamma(\phi'_5\to b+\bar{b})&=&2.68\times 10^{-4}\Gamma_{2d} \\
\Gamma(\phi'_5\to\tau+\bar{\tau})&=&4.90\times 10^{-5}\Gamma_{2d},\\
\Gamma(\rho'_6\to c+\bar{c})&=&1.91\times 10^{-7}\Gamma_{2d} ,\\
\Gamma(\rho'_6\to b+\bar{b})&=&4.96\times 10^{-6}\Gamma_{2d} ,\\
\Gamma(\rho'_6\to\tau+\bar{\tau})&=&6.89\times 10^{-8}\Gamma_{2d},
}
where $\Gamma_{2d}$ is 2-body decay width of scalar.
The interactions with the neutralinos $\eta'_1,\eta'_4,$
\eqn{
{\cal L}&\supset&-\lambda_1\L\{S_3[(h^U_1)^0(h^D_1)^0+(h^U_2)^0(h^D_2)^0]
+s_3[(H^U_1)^0(h^D_1)^0+(H^U_2)^0(h^D_2)^0] \R. \no \\
&+&\L. s_3[(h^U_1)^0(H^D_1)^0+(h^U_2)^0(H^D_2)^0] \R\} \no \\
&-&\lambda_3[S_3(h^U_3)^0(h^D_3)^0+s_3(H^U_3)^0(h^D_3)^0+s_3(h^U_3)^0(H^D_3)^0] \no \\
&-&\lambda_4\L\{(H^U_3)^0[s_1(h^D_1)^0+s_2(h^D_2)^0]
+(h^U_3)^0[S_1(h^D_1)^0+S_2(h^D_2)^0] \R. \no \\
&+&\L. (h^U_3)^0[s_1(H^D_1)^0+s_2(H^D_2)^0] \R\} \no \\
&-&\lambda_5\L\{(H^D_3)^0[s_1(h^U_1)^0+s_2(h^U_2)^0]
+(h^D_3)^0[S_1(h^U_1)^0+S_2(h^U_2)^0] \R. \no \\
&+&\L. (h^D_3)^0[s_1(H^U_1)^0+s_2(H^U_2)^0] \R\} \no \\
&-&i\sqrt{2}\sum_i[(H^U_i)^0]^\dagger
\L[-\frac12 g_2\lambda^3_2+\frac12g_Y\lambda_Y-2g_x\lambda_X\R](h^U_i)^0 \no \\
&-&i\sqrt{2}\sum_i[(H^D_i)^0]^\dagger
\L[\frac12 g_2\lambda^3_2-\frac12g_Y\lambda_Y-3g_x\lambda_X\R](h^D_i)^0 \no \\
&-&i\sqrt{2}\sum_iS^\dagger_i[5g_x\lambda_X]s_i+h.c. \no \\
&=&-(0.0620\phi'_4+0.127\phi'_5 +0.0299i\rho'_6)\eta'_4\eta'_4\no \\
&-&(-0.0716\phi'_4+0.117\phi'_5 + 0.0142i\rho'_6)\eta'_1\eta'_1+h.c.
}
give
\eqn{
\Gamma(\phi'_4\to \eta+\eta) &=&208\times 10^{-4}\Gamma_{2d} ,\\
\Gamma(\phi'_5\to \eta+\eta) &=&864\times 10^{-4}\Gamma_{2d} ,\\
\Gamma(\rho'_6\to \eta+\eta)&=& 3179 \times 10^{-6}\Gamma_{2d},
}
which dominate the decay widths of $\phi'_4,\phi'_5, \rho'_6$.

The decay widths of $(H^\pm_3)'$ due to Yukawa interactions
\eqn{
{\cal L}&\supset&-Y_c(H^U_3)^+sc^c-Y^E_1[(H^D_1)^-e^c\nu_1+(H^D_2)^-e^c_1\nu_2]+h.c. 
}
are given by
\eqn{
\Gamma((H^-_3)'\to s+\bar{c})&=&1.95\times 10^{-7}\Gamma_{2d} ,\\
\Gamma((H^-_3)'\to \tau+\bar{\nu}_\tau)&=&2.48\times 10^{-6}\Gamma_{2d}.
}
From these estimations, $(H^\pm_3)'$ decay gives dominant contribution to anti-proton flux.

\begin{figure}[ht]
\unitlength=1mm \hspace{1cm}
\begin{picture}(70,70)
\includegraphics[height=6cm,width=10cm]{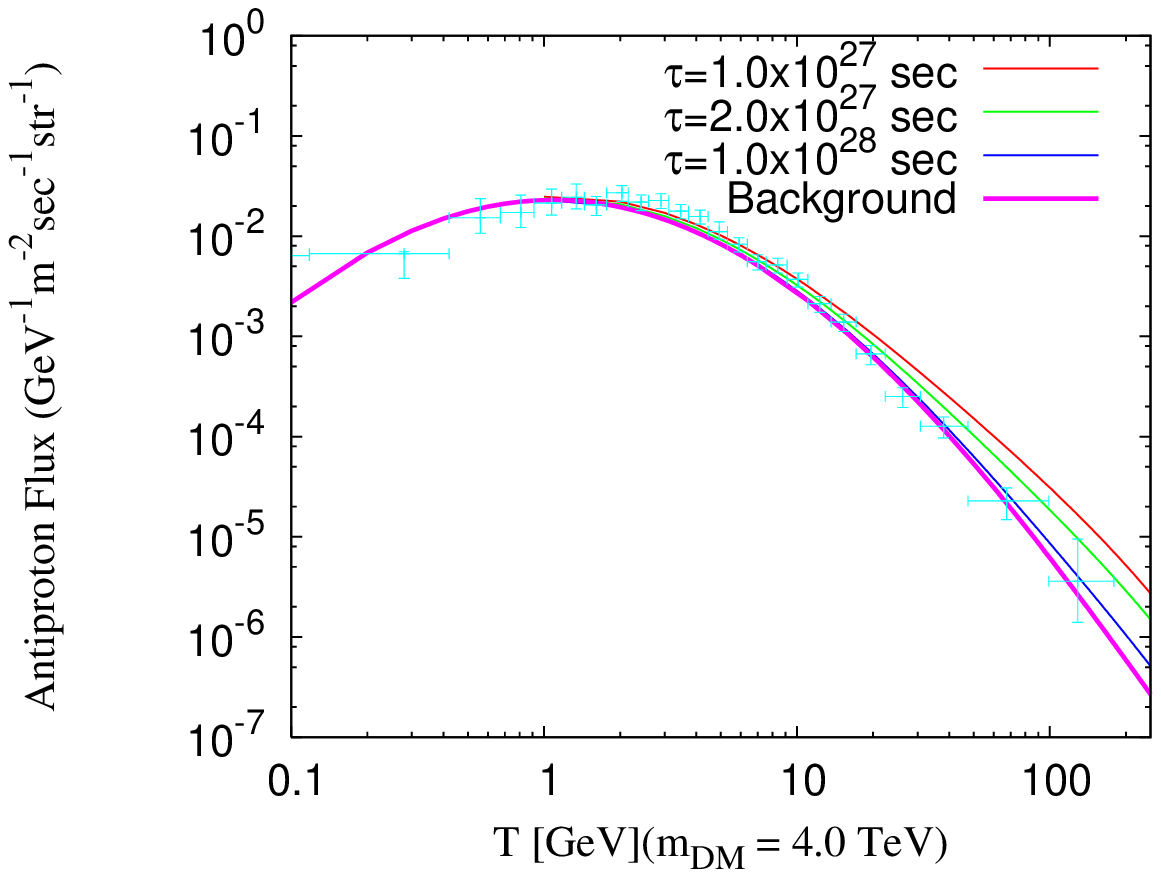}
\end{picture}
\caption{   }
\label{proton}
\end{figure}

As one charged lepton emission from LF decay accommodates one charged Higgs emission
at even rate of $(H^\pm_1)'$ and $(H^\pm_3)'$, the quark flux is estimated as
\eqn{
\tau_{\mbox{quark}}=\L[\frac12\L(0+\frac{0.195}{2.48+0.195}\R)\R]^{-1}
\tau_{\mbox{eff}}=27.4\tau_{\mbox{eff}}.
}
For each parameter sets, we get
\eqn{
\mbox{A}&:&\tau_{\mbox{quark}}=1.6\times 10^{27}\mbox{sec}, \\
\mbox{B}&:&\tau_{\mbox{quark}}=1.9\times 10^{27}\mbox{sec},
}
from which the spectrum of anti-proton flux is given in Fig. \ref{proton}.
There is no inconsistency in anti-proton flux.

\subsection{Relic Abundance of LSP}

Finally we estimate the relic abundance of LSP.
The interactions between $\eta'_1,\eta'_4$ and $Z$ are given by
\eqn{
{\cal L}&\supset&\frac12\bar{\psi}_1\gamma^\mu(-\pl_\mu-iG_1Z_\mu\gamma_5)\psi_1
+\frac12\bar{\psi}_4\gamma^\mu(-\pl_\mu-iG_4Z_\mu\gamma_5)\psi_4 \no \\
&+& iG(f_L)\bar{\psi}_f\gamma^\mu Z_\mu P_L\psi_f
+iG(f_R)\bar{\psi}_f\gamma^\mu Z_\mu P_R\psi_f,
}
where
\eqn{
&&G_1=0.00823 ,\quad
G_4=0.0119 ,\\
&&G(e_L)=-\frac{g^2_Y}{\sqrt{g^2_Y+g^2_2}}=-0.172,\quad 
G(e_R)=\frac{-g^2_Y+g^2_2}{2\sqrt{g^2_Y+g^2_2}}=0.200 ,\\
&&G(\nu_L)=-\frac{\sqrt{g^2_Y+g^2_2}}{2}=-0.372 ,\\
&&G(u_L)=\frac{2g^2_Y}{3\sqrt{g^2_Y+g^2_2}}=0.114,\quad
G(u_R)=\frac{g^2_Y-3g^2_2}{6\sqrt{g^2_Y+g^2_2}}=-0.257 ,\\
&&G(d_L)=-\frac{g^2_Y}{3\sqrt{g^2_Y+g^2_2}}=-0.057,\quad
G(d_R)=\frac{g^2_Y+3g^2_2}{6\sqrt{g^2_Y+g^2_2}}=0.315.
} 
These interactions give dominant contribution to annihilation of $\eta'_1,\eta'_4$.
As $G_i \ll G(\nu_L)$, the contributions $Z\to \eta\eta$ to Z-decay width is negligible.
Therefore LEP bound $m\geq 46 \mbox{GeV}$ is not imposed on $\eta'_1,\eta'_4$.
The relic abundance of the neutralino is calculated by the formula 
\eqn{
x_F&=&\ln\frac{0.0955m_Pm_i(a+6b/x_F)}{(g_*x_F)^\frac12}\quad (i=1,4), \\
\Omega h^2&=&\frac{8.76\times 10^{-11}g^{-\frac12}_*x_F}{(a+3b/x_F)\mbox{GeV}^2},
}
where $m_1=41.92\mbox{GeV}$, $m_4=44.55\mbox{GeV}$,
\eqn{
a_{f,i}&=&\frac{2c_f}{\pi}\L[\frac{m_f(G_i/2)}{4m^2_i-M^2_Z}
(G(f_L)-G(f_R))\R]^2\L(1-\frac{m^2_f}{m^2_i}\R)^\frac12 , \\
b_{f,i}&=&\frac16\L(-\frac92+\frac34\frac{m^2_f}{m^2_i-m^2_f}\R)a_{f,i} \no \\
&+&\frac{c_f}{3\pi}\L[\frac{m_i(G_i/2)}{4m^2_i-m^2_Z}\R]^2
\L[G^2(f_L)+G^2(f_R)\R]\L(4+\frac{2m^2_f}{m^2_i}\R)
\L(1-\frac{m^2_f}{m^2_i}\R)^\frac12 , 
}
$c_f$ is color factor such as $c_f=1$ for $SU_c(3)$ singlet, $c_f=3$ for triplet, and $m_Z=91.2$ GeV.
For the approximation $m_f/m_i=0$, these coefficients are given by
\eqn{
a_1&=&a_{b,1}+a_{\tau,1}=5.01\times 10^{-11}\mbox{GeV}^{-2} , \\
a_4&=&a_{b,4}+a_{\tau,4}=1.22\times 10^{-9}\mbox{GeV}^{-2} ,\\
b_1&=&1.53\times 10^{-8} \mbox{GeV}^{-2} ,\\
b_4&=&4.23\times 10^{-7} \mbox{GeV}^{-2}.
}
The relic abundance of $\eta'_1$ is given by
\eqn{
g_*&=&75.75 ,\no \\
x_F&=&22.32 ,\no \\
T_F&=&1.88 \mbox{ GeV} ,\no \\
\Omega_1h^2&=&0.106,
}
and that of $\eta'_4$ is given by
\eqn{
g_*&=&72.25 ,\no \\
x_F&=&25.52 ,\no \\
T_F&=&1.75 \mbox{ GeV} ,\no \\
\Omega_4h^2&=&0.0052.
}
As $\eta'_4$ is converted into $\eta'_1$, relic abundance of LSP is given by
\eqn{
(\Omega_{CDM}h^2)=\Omega_1h^2+\Omega_4h^2=0.111,
}
which realizes density parameter of dark matter.

\section{Conclusion}

We have considered dark matter based on
$S_4\times Z_2$ flavor symmetric extra U(1) model.
The results are as follows.
There exists appropriate parameter set to realize
relic abundance of dark matter and positron flux observed by PAMELA at the same time.
The dominant component of dark matter is LSP and the origin of positron flux is
given by the decay of LF which generates the mass of RHN.
There is deep connection between PAMELA observation and neutrino mass.
The long life time of LF results in large RHN mass and
the spectrum of positron flux depend on Majorana phase $\phi$ in $V_{MNS}$.
Therefore, cosmic-ray observation gives new information about the
structure of neutrino mass matrix.

From the fact that there is no excess of anti-proton flux in cosmic-ray,
we can guess about the particle spectrum.
As sfermions can decay into quarks, weak boson and charginos,
LF must not decay into those particles, which sugests sfermions are heavier than 4TeV.
Although this is also favorable from the viewpoint of the FCNC constraints,
experimental verification becomes difficult.
However experimantal verification of our scenario is not imposssible.
From the fact that Higgs does not decay into weak boson or charginos, 
we can expect that Higgs boson is light and degenerated, 
therefore the examination of the mass spectrum of Higgs boson is possible.

\appendix

\section{Mass matrices}

{\bf Neutral CP even Higgs boson}
\eqn{
H^U_a&\supset&\frac{\phi_{U,a}}{\sqrt{2}},\quad H^D_a\supset\frac{\phi_{D,a}}{\sqrt{2}},\quad
S_a\supset\frac{\phi_{S,a}}{\sqrt{2}}\quad (a=1,2,3), \no \\
-{\cal L}&\supset&\frac12\phi_i M^2_{ij}\phi_j,\quad \phi_i=\3tvec{\phi_{U,a}}{\phi_{D,a}}{\phi_{S,a}},
\quad (i,j=1,2,\cdots ,9), \no
}
\eqn{
M^2_{1,1}&=&M^2_{2,2}
=\sqrt{2}m^2_{BU}(v'_u/v_u)+\lambda_1A_1v'_s(v_d/v_u)+\lambda_5A_5v_s(v'_d/v_u)
-\lambda^2_1v^2_d/2-\lambda^2_5v^2_s/2 \no \\
&-&\L[(\lambda_1\lambda_4+\lambda_3\lambda_5)v_sv'_s
+(\lambda_4\lambda_5+\lambda_1\lambda_3)v'_dv_d\R](v'_u/v_u)
+\L[\frac14(g^2_Y+g^2_2)+4g^2_x\R]v^2_u ,\no \\
M^2_{1,2}&=&\lambda^2_5v^2_s/2+\lambda^2_1v^2_d/2
+\L[\frac14(g^2_Y+g^2_2)+4g^2_x\R]v^2_u ,\no \\
M^2_{1,3}&=&M^2_{2,3}
=-m^2_{BU}+[(\lambda_1\lambda_4+\lambda_3\lambda_5)v'_sv_s
+(\lambda_4\lambda_5+\lambda_1\lambda_3)v_dv'_d]/\sqrt{2} \no \\
&+&\sqrt{2}\L[\frac14(g^2_Y+g^2_2)+4g^2_x\R]v_uv'_u ,\no \\
M^2_{3,3}
&=&\sqrt{2}m^2_{BU}(v_u/v'_u)+\lambda_3A_3v'_s(v'_d/v'_u)+\lambda_4A_4v_s(v_d/v'_u) \no \\
&-&\L[(\lambda_1\lambda_4+\lambda_3\lambda_5)v_sv'_s
+(\lambda_4\lambda_5+\lambda_1\lambda_3)v'_dv_d\R](v_u/v'_u) 
+2\L[\frac14(g^2_Y+g^2_2)+4g^2_x\R](v'_u)^2 ,\no \\
M^2_{4,4}&=&M^2_{5,5}
=\sqrt{2}m^2_{BD}(v'_d/v_d)+\lambda_1A_1v'_s(v_u/v_d)+\lambda_4A_4v_s(v'_u/v_d)
-\lambda^2_1v^2_u/2-\lambda^2_4v^2_s/2 \no \\
&-&\L[(\lambda_1\lambda_5+\lambda_3\lambda_4)v_sv'_s
+(\lambda_4\lambda_5+\lambda_1\lambda_3)v'_uv_u\R](v'_d/v_d)
+v^2_d\L[\frac14(g^2_Y+g^2_2)+9g^2_x\R] ,\no \\
M^2_{4,5}&=&\lambda^2_4v^2_s/2+\lambda^2_1v^2_u/2
+\L[\frac14(g^2_Y+g^2_2)+9g^2_x\R]v^2_d ,\no \\
M^2_{4,6}&=&M^2_{5,6}
=-m^2_{BD}+[(\lambda_1\lambda_5+\lambda_3\lambda_4)v'_sv_s
+(\lambda_4\lambda_5+\lambda_1\lambda_3)v_uv'_u]/\sqrt{2} \no \\
&+&\sqrt{2}\L[\frac14(g^2_Y+g^2_2)+9g^2_x\R]v_dv'_d ,\no \\
M^2_{6,6}&=&\sqrt{2}m^2_{BD}(v_d/v'_d)
+\lambda_3A_3v'_s(v'_u/v'_d)+\lambda_5A_5v_s(v_u/v'_d) \no \\
&-&\L[(\lambda_1\lambda_5+\lambda_3\lambda_4)v_sv'_s
+(\lambda_4\lambda_5+\lambda_1\lambda_3)v'_uv_u\R](v_d/v'_d)
+2\L[\frac14(g^2_Y+g^2_2)+9g^2_x\R](v'_d)^2 ,\no \\
M^2_{7,7}&=&M^2_{8,8}
=\sqrt{2}m^2_{BS}(v'_s/v_s)+\lambda_4A_4(v'_uv_d/v_s)+\lambda_5A_5(v_uv'_d/v_s)
-\lambda^2_4v^2_d/2-\lambda^2_5v^2_u/2 \no \\
&-&\L[(\lambda_1\lambda_5+\lambda_3\lambda_4)v_dv'_d
+(\lambda_1\lambda_4+\lambda_3\lambda_5)v'_uv_u\R](v'_s/v_s)
+\L[25g^2_x\R]v^2_s ,\no \\
M^2_{7,8}&=&\lambda^2_4v^2_d/2+\lambda^2_5v^2_u/2
+\L[25g^2_x\R]v^2_s ,\no \\
M^2_{7,9}&=&M^2_{8,9}
=-m^2_{BS}+[(\lambda_1\lambda_5+\lambda_3\lambda_4)v'_dv_d
+(\lambda_1\lambda_4+\lambda_3\lambda_5)v_uv'_u]/\sqrt{2}
+\sqrt{2}\L[25g^2_x\R]v_sv'_s ,\no \\
M^2_{9,9}
&=&\sqrt{2}m^2_{BS}(v_s/v'_s)+\lambda_1A_1(v_uv_d/v'_s)+\lambda_3A_3(v'_uv'_d/v'_s) \no \\
&-&\L[(\lambda_1\lambda_5+\lambda_3\lambda_4)v_dv'_d
+(\lambda_1\lambda_4+\lambda_3\lambda_5)v'_uv_u\R](v_s/v'_s)
+2\L[25g^2_x\R](v'_s)^2,\no \\
M^2_{1,4}&=&M^2_{2,5}
=-\lambda_1A_1v'_s+3\lambda^2_1v_uv_d/2+(\lambda_4\lambda_5+\lambda_1\lambda_3)v'_uv'_d
+\L[-\frac14(g^2_Y+g^2_2)+6g^2_x\R]v_uv_d ,\no \\
M^2_{1,5}&=&M^2_{2,4}=\lambda^2_1v_uv_d/2
+\L[-\frac14(g^2_Y+g^2_2)+6g^2_x\R]v_uv_d ,\no \\
M^2_{1,6}&=&M^2_{2,6}
=-\lambda_5A_5v_s/\sqrt{2}+\sqrt{2}\lambda^2_5v_uv'_d
+(\lambda_4\lambda_5+\lambda_1\lambda_3)v'_uv_d/\sqrt{2} \no \\
&+&\sqrt{2}\L[-\frac14(g^2_Y+g^2_2)+6g^2_x\R]v_uv'_d ,\no \\
M^2_{3,4}&=&M^2_{3,5}
=-\lambda_4A_4v_s/\sqrt{2}+\sqrt{2}\lambda^2_4v'_uv_d
+(\lambda_4\lambda_5+\lambda_1\lambda_3)v_uv'_d/\sqrt{2} \no \\
&+&\sqrt{2}\L[-\frac14(g^2_Y+g^2_2)+6g^2_x\R]v'_uv_d ,\no \\
M^2_{3,6}&=&-\lambda_3A_3v'_s
+2\lambda^2_3v'_uv'_d+(\lambda_1\lambda_3+\lambda_4\lambda_5)v_uv_d
+2\L[-\frac14(g^2_Y+g^2_2)+6g^2_x\R]v'_uv'_d ,\no \\
M^2_{1,7}&=&M^2_{2,8}
=-\lambda_5A_5v'_d
+3\lambda^2_5v_uv_s/2+(\lambda_1\lambda_4+\lambda_3\lambda_5)v'_uv'_s
+\L[-10g^2_x\R]v_uv_s ,\no \\
M^2_{1,8}&=&M^2_{2,7}
=\lambda^2_5v_uv_s/2
+\L[-10g^2_x\R]v_uv_s ,\no \\
M^2_{1,9}&=&M^2_{2,9}
=-\lambda_1A_1v_d/\sqrt{2}+\sqrt{2}\lambda^2_1v_uv'_s
+(\lambda_1\lambda_4+\lambda_3\lambda_5)v'_uv_s/\sqrt{2}
+\sqrt{2}\L[-10g^2_x\R]v_uv'_s ,\no \\
M^2_{3,7}&=&M^2_{3,8}
=-\lambda_4A_4v_d/\sqrt{2}+\sqrt{2}\lambda^2_4v'_uv_s
+(\lambda_1\lambda_4+\lambda_3\lambda_5)v_uv'_s/\sqrt{2}
+\sqrt{2}\L[-10g^2_x\R]v'_uv_s ,\no \\
M^2_{3,9}&=&-\lambda_3A_3v'_d
+2\lambda^2_3v'_uv'_s+(\lambda_1\lambda_4+\lambda_3\lambda_5)v_uv_s
+2\L[-10g^2_x\R]v'_uv'_s ,\no \\
M^2_{4,7}&=&M^2_{5,8}
=-\lambda_4A_4v'_u+3\lambda^2_4v_dv_s/2
+(\lambda_1\lambda_5+\lambda_3\lambda_4)v'_dv'_s
+\L[-15g^2_x\R]v_dv_s ,\no \\
M^2_{4,8}&=&M^2_{5,7}
=\lambda^2_4v_dv_s/2
+\L[-15g^2_x\R]v_dv_s ,\no \\
M^2_{4,9}&=&M^2_{5,9}
=-\lambda_1A_1v_u/\sqrt{2}+\sqrt{2}\lambda^2_1v_dv'_s
+(\lambda_1\lambda_5+\lambda_3\lambda_4)v'_dv_s/\sqrt{2}
+\sqrt{2}\L[-15g^2_x\R]v_dv'_s ,\no \\
M^2_{6,7}&=&M^2_{6,8}
=-\lambda_5A_5v_u/\sqrt{2}+\sqrt{2}\lambda^2_5v'_dv_s
+(\lambda_1\lambda_5+\lambda_3\lambda_4)v_dv'_s/\sqrt{2}
+\sqrt{2}\L[-15g^2_x\R]v'_dv_s ,\no \\
M^2_{6,9}&=&-\lambda_3A_3v'_u+2\lambda^2_3v'_dv'_s
+(\lambda_1\lambda_5+\lambda_3\lambda_4)v_dv_s
+2\L[-15g^2_x\R]v'_dv'_s.
}

{\bf Neutral CP odd Higgs boson}
\eqn{
H^U_a&\supset&\frac{i\rho_{U,a}}{\sqrt{2}},\quad H^D_a\supset\frac{i\rho_{D,a}}{\sqrt{2}},\quad
S_a\supset\frac{i\rho_{S,a}}{\sqrt{2}}\quad (a=1,2,3), \no \\
-{\cal L}&\supset&\frac12\rho_i M^2_{ij}\rho_j,\quad \rho_i=\3tvec{\rho_{U,a}}{\rho_{D,a}}{\rho_{S,a}},
\quad (i,j=1,2,\cdots ,9), \no 
}
\eqn{
M^2_{1,1}&=&M^2_{2,2}
=\sqrt{2}m^2_{BU}(v'_u/v_u)+\lambda_1A_1v'_s(v_d/v_u)+\lambda_5A_5v_s(v'_d/v_u)
-\lambda^2_1v^2_d/2-\lambda^2_5v^2_s/2 \no \\
&-&\L[(\lambda_1\lambda_4+\lambda_3\lambda_5)v_sv'_s
+(\lambda_4\lambda_5+\lambda_1\lambda_3)v'_dv_d\R](v'_u/v_u) ,\no \\
M^2_{1,2}&=&\lambda^2_5v^2_s/2+\lambda^2_1v^2_d/2 ,\no \\
M^2_{1,3}&=&M^2_{2,3}
=-m^2_{BU}+[(\lambda_1\lambda_4+\lambda_3\lambda_5)v'_sv_s
+(\lambda_4\lambda_5+\lambda_1\lambda_3)v_dv'_d]/\sqrt{2} ,\no \\
M^2_{3,3}
&=&\sqrt{2}m^2_{BU}(v_u/v'_u)+\lambda_3A_3v'_s(v'_d/v'_u)+\lambda_4A_4v_s(v_d/v'_u) \no \\
&-&\L[(\lambda_1\lambda_4+\lambda_3\lambda_5)v_sv'_s
+(\lambda_4\lambda_5+\lambda_1\lambda_3)v'_dv_d\R](v_u/v'_u) ,\no \\
M^2_{4,4}&=&M^2_{5,5}
=\sqrt{2}m^2_{BD}(v'_d/v_d)+\lambda_1A_1v'_s(v_u/v_d)+\lambda_4A_4v_s(v'_u/v_d)
-\lambda^2_1v^2_u/2-\lambda^2_4v^2_s/2 \no \\
&-&\L[(\lambda_1\lambda_5+\lambda_3\lambda_4)v_sv'_s
+(\lambda_4\lambda_5+\lambda_1\lambda_3)v'_uv_u\R](v'_d/v_d) ,\no \\
M^2_{4,5}&=&\lambda^2_4v^2_s/2+\lambda^2_1v^2_u/2 ,\no \\
M^2_{4,6}&=&M^2_{5,6}
=-m^2_{BD}+[(\lambda_1\lambda_5+\lambda_3\lambda_4)v'_sv_s
+(\lambda_4\lambda_5+\lambda_1\lambda_3)v_uv'_u]/\sqrt{2} ,\no  \\
M^2_{6,6}
&=&\sqrt{2}m^2_{BD}(v_d/v'_d)+\lambda_3A_3v'_s(v'_u/v'_d)+\lambda_5A_5v_s(v_u/v'_d) \no \\
&-&\L[(\lambda_1\lambda_5+\lambda_3\lambda_4)v_sv'_s
+(\lambda_4\lambda_5+\lambda_1\lambda_3)v'_uv_u\R](v_d/v'_d) ,\no \\
M^2_{7,7}&=&M^2_{8,8}
=\sqrt{2}m^2_{BS}(v'_s/v_s)+\lambda_4A_4(v'_uv_d/v_s)+\lambda_5A_5(v_uv'_d/v_s)
-\lambda^2_4v^2_d/2-\lambda^2_5v^2_u/2 \no \\
&-&\L[(\lambda_1\lambda_5+\lambda_3\lambda_4)v_dv'_d
+(\lambda_1\lambda_4+\lambda_3\lambda_5)v'_uv_u\R](v'_s/v_s) ,\no \\
M^2_{7,8}&=&\lambda^2_4v^2_d/2+\lambda^2_5v^2_u/2 ,\no \\
M^2_{7,9}&=&M^2_{8,9}
=-m^2_{BS}+[(\lambda_1\lambda_5+\lambda_3\lambda_4)v'_dv_d
+(\lambda_1\lambda_4+\lambda_3\lambda_5)v_uv'_u]/\sqrt{2} ,\no \\
M^2_{9,9}
&=&\sqrt{2}m^2_{BS}(v_s/v'_s)+\lambda_1A_1(v_uv_d/v'_s)+\lambda_3A_3(v'_uv'_d/v'_s) \no \\
&-&\L[(\lambda_1\lambda_5+\lambda_3\lambda_4)v_dv'_d
+(\lambda_1\lambda_4+\lambda_3\lambda_5)v'_uv_u\R](v_s/v'_s) ,\no \\
M^2_{1,4}&=&M^2_{2,5}
=\lambda_1A_1v'_s+(\lambda_4\lambda_5-\lambda_1\lambda_3)v'_uv'_d
-\lambda^2_1v_uv_d/2 ,\no \\
M^2_{1,5}&=&M^2_{2,4}
=\lambda^2_1v_uv_d/2 ,\no \\
M^2_{1,6}&=&M^2_{2,6}
=\lambda_5A_5v_s/\sqrt{2}
+(\lambda_1\lambda_3-\lambda_4\lambda_5)v'_uv_d/\sqrt{2} ,\no \\
M^2_{3,4}&=&M^2_{3,5}
=\lambda_4A_4v_s/\sqrt{2}
+(\lambda_1\lambda_3-\lambda_4\lambda_5)v_uv'_d/\sqrt{2} ,\no \\
M^2_{3,6}&=&\lambda_3A_3v'_s
+(\lambda_4\lambda_5-\lambda_1\lambda_3)v_uv_d ,\no \\
M^2_{1,7}&=&M^2_{2,8}
=\lambda_5A_5v'_d+(\lambda_1\lambda_4-\lambda_3\lambda_5)v'_uv'_s
-\lambda^2_5v_uv_s/2 ,\no \\
M^2_{1,8}&=&M^2_{2,7}
=\lambda^2_5v_uv_s/2 ,\no \\
M^2_{1,9}&=&M^2_{2,9}
=\lambda_1A_1v_d/\sqrt{2}
+(\lambda_3\lambda_5-\lambda_1\lambda_4)v'_uv_s/\sqrt{2} ,\no \\
M^2_{3,7}&=&M^2_{3,8}
=\lambda_4A_4v_d/\sqrt{2}
+(\lambda_3\lambda_5-\lambda_1\lambda_4)v_uv'_s/\sqrt{2} ,\no \\
M^2_{3,9}&=&\lambda_3A_3v'_d
+(\lambda_1\lambda_4-\lambda_3\lambda_5)v_uv_s ,\no \\
M^2_{4,7}&=&M^2_{5,8}
=\lambda_4A_4v'_u+(\lambda_1\lambda_5-\lambda_3\lambda_4)v'_dv'_s
-\lambda^2_4v_dv_s/2 ,\no \\
M^2_{4,8}&=&M^2_{5,7}
=\lambda^2_4v_dv_s/2 ,\no \\
M^2_{4,9}&=&M^2_{5,9}
=\lambda_1A_1v_u/\sqrt{2}
+(\lambda_3\lambda_4-\lambda_1\lambda_5)v'_dv_s/\sqrt{2} ,\no \\
M^2_{6,7}&=&M^2_{6,8}
=\lambda_5A_5v_u/\sqrt{2}
+(\lambda_3\lambda_4-\lambda_1\lambda_5)v_dv'_s/\sqrt{2} ,\no \\
M^2_{6,9}&=&\lambda_3A_3v'_u
+(\lambda_1\lambda_5-\lambda_3\lambda_4)v_dv_s.
}

{\bf Charged Higgs boson}
\eqn{
&&H^U_a\supset H^+_a,\quad H^D_a\supset H^-_{a+3}\quad (a=1,2,3) ,\no \\
&&-{\cal L}\supset H^+_iM^2_{ij}H^-_j,\quad H^+_i=\2tvec{H^+_a}{(H^-_{a+3})^\dagger}\quad 
(i,j=1,2,\cdots,6) ,\no
}

\eqn{
M^2_{1,1}&=&\sqrt{2}m^2_{BU}(v'_u/v_u)+\lambda_1A_1v'_s(v_d/v_u)+\lambda_5A_5v_s(v'_d/v_u)
-\lambda^2_1v^2_d-\lambda^2_5[v^2_s/2+(v'_d)^2] \no \\
&-&\L[(\lambda_1\lambda_4+\lambda_3\lambda_5)v_sv'_s
+(\lambda_4\lambda_5+\lambda_1\lambda_3)v'_dv_d\R](v'_u/v_u) \no \\
&-&\frac12g^2_2[v^2_u+(v'_u)^2-v^2_d-(v'_d)^2]+\frac14g^2_2v^2_u ,\no \\
M^2_{1,2}&=&\lambda^2_5v^2_s/2+\frac14g^2_2v^2_u ,\no \\
M^2_{1,3}&=&M^2_{2,3}
=-m^2_{BU}+(\lambda_1\lambda_4+\lambda_3\lambda_5)v_sv'_s/\sqrt{2}
+\frac{1}{2\sqrt{2}}g^2_2v'_uv_u ,\no \\
M^2_{3,3}
&=&\sqrt{2}m^2_{BU}(v_u/v'_u)+\lambda_3A_3v'_s(v'_d/v'_u)+\lambda_4A_4v_s(v_d/v'_u) 
-\lambda^2_4v^2_d-\lambda^2_3(v'_d)^2 \no \\
&-&\L[(\lambda_1\lambda_4+\lambda_3\lambda_5)v_sv'_s
+(\lambda_4\lambda_5+\lambda_1\lambda_3)v'_dv_d\R](v_u/v'_u) \no \\
&-&\frac12g^2_2[v^2_u+(v'_u)^2-v^2_d-(v'_d)^2]+\frac12g^2_2(v'_u)^2 ,\no \\
M^2_{4,4}&=&M^2_{5,5}
=\sqrt{2}m^2_{BD}(v'_d/v_d)+\lambda_1A_1v'_s(v_u/v_d)+\lambda_4A_4v_s(v'_u/v_d) \no \\
&-&\lambda^2_1v^2_u-\lambda^2_4[v^2_s/2+(v'_u)^2] 
-\L[(\lambda_1\lambda_5+\lambda_3\lambda_4)v_sv'_s
+(\lambda_4\lambda_5+\lambda_1\lambda_3)v'_uv_u\R](v'_d/v_d) \no \\
&+&\frac12g^2_2[v^2_u+(v'_u)^2-v^2_d-(v'_d)^2]+\frac14g^2_2v^2_d ,\no \\
M^2_{4,5}&=&\lambda^2_4v^2_s/2+\frac14g^2_2v^2_d ,\no \\
M^2_{4,6}&=&M^2_{5,6}
=-m^2_{BD}+(\lambda_1\lambda_5+\lambda_3\lambda_4)v_sv'_s/\sqrt{2}
+\frac{1}{2\sqrt{2}}g^2_2v'_dv_d ,\no \\
M^2_{6,6}
&=&\sqrt{2}m^2_{BD}(v_d/v'_d)+\lambda_3A_3v'_s(v'_u/v'_d)+\lambda_5A_5v_s(v_u/v'_d)
-\lambda^2_5v^2_u-\lambda^2_3(v'_u)^2 \no \\
&-&\L[(\lambda_1\lambda_5+\lambda_3\lambda_4)v_sv'_s
+(\lambda_4\lambda_5+\lambda_1\lambda_3)v'_uv_u\R](v_d/v'_d)\no \\
&+&\frac12g^2_2[v^2_u+(v'_u)^2-v^2_d-(v'_d)^2]+\frac12g^2_2(v'_d)^2 ,\no \\
M^2_{1,4}&=&M^2_{2,5}
=\lambda_1A_1v'_s-\lambda_1[\lambda_1v_uv_d+\lambda_3v'_uv'_d]+\frac14g^2_2v_uv_d ,\no \\
M^2_{1,5}&=&M^2_{2,4}
=\frac14g^2_2v_uv_d ,\no \\
M^2_{1,6}&=&M^2_{2,6}
=\lambda_5A_5v_s/\sqrt{2}-\lambda_5[\lambda_4v'_uv_d+\lambda_5v_uv'_d]/\sqrt{2}
+\frac{1}{2\sqrt{2}}g^2_2v_uv'_d ,\no \\
M^2_{3,4}&=&M^2_{3,5}
=\lambda_4A_4v_s/\sqrt{2}-\lambda_4[\lambda_4v'_uv_d+\lambda_5v_uv'_d]/\sqrt{2}
+\frac{1}{2\sqrt{2}}g^2_2v'_uv_d ,\no \\
M^2_{3,6}&=&\lambda_3A_3v'_s-\lambda_3[\lambda_1v_uv_d+\lambda_3v'_uv'_d]
+\frac{1}{2}g^2_2v'_uv'_d.
}

{\bf 1-loop corrections to Higgs mass}

In order to satisfy the experimental bound for the
lightest neutral CP even Higgs boson mass, the contributions from
1-loop corrections are important \cite{1-loop}\cite{1-loop-u1}. We add 1-loop contributions
\eqn{
\Delta V&=&\frac{1}{64\pi^2}Str\L[M^4\L(\ln\frac{M^2}{\Lambda^2}-\frac32\R)\R]
}
to Higgs potentail.The dominant contributions are given by
trilinear terms $Y_tH^U_3Q_3U^c_3$ and $kS_3(g_1g^c_1+g_2g^c_2+g_3g^c_3)$.
From the mass terms of squark and scalar g-quark
\eqn{
-{\cal L}&\supset&\L[m^2_{Q_3}+(Y_t(H^U_3)^0)^2\R]|U_3|^2
+\L[m^2_{U^c_3}+(Y_t(H^U_3)^0)^2\R]|U^c_3|^2 \no \\
&+&\L[m^2_g+(kv'_s)^2\R](|g_1|^2+|g_2|^2+|g_3|^2) 
+\L[m^2_{g^c}+(kv'_s)^2\R](|g^c_1|^2+|g^c_2|^2+|g^c_3|^2)  \no \\
&+&\L[Y_tA_t(H^U_3)^0+Y_t\lambda_3v'_s((H^D_3)^0)^*\R]U_3U^c_3 \no \\
&+&\L[kA_kv'_s+k\lambda_3((H^U_3)^0(H^D_3)^0)^*\R](g_1g^c_1+g_2g^c_2+g_3g^c_3)+h.c. \no \\
&=&(U^*_3,U^c_3)
\mat2{m^2_{Q_3}+(Y_tv'_u)^2}{Y_tA_tv'_u+Y_t\lambda_3v'_sv'_d}
{Y_tA_tv'_u+Y_t\lambda_3v'_sv'_d}{m^2_{U^c_3}+(Y_tv'_u)^2}\2tvec{U_3}{(U^c_3)^*} \no \\
&+&\sum_i(g^*_i,g^c_i)
\mat2{m^2_g+(kv'_s)^2}{kA_kv'_s+k\lambda_3v'_uv'_d}
{kA_kv'_s+k\lambda_3v'_uv'_d}{m^2_{g^c}+(kv'_s)^2}\2tvec{g_i}{(g^c_i)^*},
}
mass eigenvalues are given by
\eqn{
M^2_{T,\pm}&=&\frac12\L[m^2_{Q_3}+m^2_{U^c_3}+2(Y_tv'_u)^2 
\pm \sqrt{(m^2_{Q_3}-m^2_{U^c_3})^2+4Y^2_t(A_tv'_u+\lambda_3v'_sv'_d)^2}\R], \no  \\
M^2_{g,\pm}&=&\frac12\L[m^2_g+m^2_{g^c}+2(kv'_s)^2
\pm \sqrt{(m^2_g-m^2_{g^c})^2+4k^2(A_kv'_s+\lambda_3v'_uv'_d)^2}\R].
}
For simplicity, we assume
\eqn{
m^2_{Q_3}=m^2_{U^c_3}=m^2_g=m^2_{g^c}=m^2_Q=16\mbox{TeV}^2, \quad A_t=A_k=0.0\mbox{TeV},
}
then Eq.(118) is rewritten by
\eqn{
M^2_{T,\pm}&=&m^2_Q+(Y_tv'_u)^2 \pm Y_t\lambda_3v'_sv'_d ,\no  \\
M^2_{g,\pm}&=&m^2_Q+(kv'_s)^2 \pm k\lambda_3v'_uv'_d.
}
As potential minimum condition is modified as
\eqn{
\frac{\pl (V+\Delta V)}{\pl X}=0, \quad X=(v'_u,v'_d),
}
and we must add the terms
\eqn{
\Delta M^2_{3,3}&=&\frac12\frac{\pl^2\Delta V}{\pl (v'_u)^2}
-\frac{1}{2(v'_u)}\frac{\pl \Delta V}{\pl (v'_u)} ,\no \\
\Delta M^2_{6,6}&=&\frac12\frac{\pl^2\Delta V}{\pl (v'_d)^2}
-\frac{1}{2(v'_d)}\frac{\pl \Delta V}{\pl (v'_d)} ,\no \\
\Delta M^2_{3,6}&=&\frac12\frac{\pl^2\Delta V}{\pl(v'_u)\pl(v'_d)},
}
to the neutral CP-even Higgs boson mass. We fix renormalization point as
\eqn{
\Lambda=4\mbox{TeV}.
}

{\bf Chargino}
\eqn{
{\cal L}&\supset&\chi^+_iM_{ij}\chi^-_j+h.c. ,\no \\
&&h^U_a=\2tvec{h^+_u}{h^0_u}_a,\quad h^D_a=\2tvec{h^0_d}{h^-_d}_a\quad (a=1,2,3) ,\no \\
&&\chi^+_i=\2tvec{(h^+_u)_a}{-iw^+},\quad \chi^-_i=\2tvec{(h^-_d)_a}{-iw^-},\quad
w^\pm=\frac{\lambda^1_2\mp i\lambda^2_2}{\sqrt{2}} \quad (i,j=1,2,3,4) ,\no \\
M&=&\L(
\begin{array}{cccc}
\lambda_1v'_s         & 0                   &\lambda_5v_s/\sqrt{2}&g_2v_u/\sqrt{2}\\
0                     &\lambda_1v'_s        &\lambda_5v_s/\sqrt{2}&g_2v_u/\sqrt{2}\\
\lambda_4v_s/\sqrt{2} &\lambda_4v_s/\sqrt{2}&\lambda_3v'_s        &g_2v'_u\\
g_2v_d/\sqrt{2}       &g_2v_d/\sqrt{2}      &g_2v'_d              &M_2 \\
\end{array}
\R).
}

{\bf Neutralino}
\eqn{
&&{\cal L}\supset -\frac12\eta_i M_{ij}\eta_j+h.c. ,\no \\
&&\eta=\L(
\begin{array}{c}
(h^0_u)_a\\
(h^0_d)_a\\
s_a\\
i\lambda \\
\end{array}
\R),\quad
\lambda=\3tvec{\lambda_Y}{\lambda^3_2}{\lambda_X} \quad (a=1,2,3;\quad i,j=1,2,\cdots,12) ,\no \\
&&M=\L(
\begin{array}{cccc}
0            &M_{ud}       &M_{us}       &M_{u\lambda}\\
M_{du}       &0            &M_{ds}       &M_{d\lambda}\\
M_{su}       &M_{sd}       &0            &M_{s\lambda}\\
M_{\lambda u}&M_{\lambda d}&M_{\lambda s}&M_{\lambda\lambda} \no \\
\end{array}
\R),
}
\eqn{
M_{ud}&=&\Mat3{\lambda_1v'_s}{0}{\lambda_5v_s/\sqrt{2}}
{0}{\lambda_1v'_s}{\lambda_5v_s/\sqrt{2}}
{\lambda_4v_s/\sqrt{2}}{\lambda_4v_s/\sqrt{2}}{\lambda_3v'_s}=M^T_{du} ,\no \\
M_{us}&=&\Mat3{\lambda_5v'_d}{0}{\lambda_1v_d/\sqrt{2}}
{0}{\lambda_5v'_d}{\lambda_1v_d/\sqrt{2}}
{\lambda_4v_d/\sqrt{2}}{\lambda_4v_d/\sqrt{2}}{\lambda_3v'_d}=M^T_{su} ,\no \\
M_{ds}&=&\Mat3{\lambda_4v'_u}{0}{\lambda_1v_u/\sqrt{2}}
{0}{\lambda_4v'_u}{\lambda_1v_u/\sqrt{2}}
{\lambda_5v_u/\sqrt{2}}{\lambda_5v_u/\sqrt{2}}{\lambda_3v'_u}=M^T_{sd} ,\no \\
M_{u\lambda}&=&\Mat3{g_Yv_u/2}{-g_2v_u/2}{-2g_xv_u}
{g_Yv_u/2}{-g_2v_u/2}{-2g_xv_u}
{g_Yv'_u/\sqrt{2}}{-g_2v'_u/\sqrt{2}}{-2\sqrt{2}g_xv'_u}=M^T_{\lambda u},\no \\
M_{d\lambda}&=&\Mat3{-g_Yv_d/2}{g_2v_d/2}{-3g_xv_d}
{-g_Yv_d/2}{g_2v_d/2}{-3g_xv_d}
{-g_Yv'_d/\sqrt{2}}{g_2v'_d/\sqrt{2}}{-3\sqrt{2}g_xv'_d}=M^T_{\lambda d},\no \\
M_{s\lambda}&=&\Mat3{0}{0}{5g_xv_s}
{0}{0}{5g_xv_s}
{0}{0}{5\sqrt{2}g_xv'_s}=M^T_{\lambda s} ,\no \\
M_{\lambda\lambda}&=&\Mat3{-M_Y}{0}{0}
{0}{-M_2}{0}
{0}{0}{-M_X}.
}

\section{Mixing matrices  and mass eigenvalues}

Mass eigenvalues ($m_i:\mbox{GeV}$) and diagnalization matrix
$U=(u_1,u_2,\cdots)$ are given as follows.

{\bf Neutral CP-even  Higgs}\\
For Higgs bosons, the diagonalization matrices are defined as 
\eqn{
(M^2)'=U^TM^2U=\mbox{diag}(m^2_1,m^2_2,\cdots).
}
\eqn{
\begin{array}{|c|c|c|c|c|c|c|c|c|c|} 
\hline
i  & 1     & 2     & 3     & 4     & 5     & 6     & 7      & 8     & 9    \\
\hline
u_i&-0.698 &-0.113 &-0.0006&-0.613 & 0.329 &-0.115 & 0.0006 & 0.003 &-0.059\\
   & 0.698 & 0.113 & 0.0006&-0.613 & 0.329 &-0.115 & 0.0006 & 0.003 &-0.059\\ 
   & 0     & 0     & 0     &-0.384 &-0.808 &-0.036 & 0.029  &-0.009 &-0.445\\
   & 0.0015&-0.005 &-0.707 &-0.0012&-0.008 & 0.006 &-0.706  &-0.0008&-0.030\\
   &-0.0015& 0.005 & 0.707 &-0.0012&-0.008 & 0.006 &-0.706  &-0.0008&-0.030\\
   & 0     & 0     & 0     &-0.273 &-0.360 &-0.031 &-0.033  & 0.054 & 0.889\\
   & 0.113 &-0.698 &-0.0056& 0.118 &-0.025 &-0.693 & 0.006  & 0.071 &-0.002\\
   &-0.113 & 0.698 & 0.0056& 0.118 &-0.025 &-0.693 & 0.006  & 0.071 &-0.002\\
   & 0     & 0     & 0     &-0.002 & 0.014 & 0.101 & 0.00009& 0.993 &-0.052\\
\hline
m_i &91.50  &537    & 2005  &121.96 &152.96 &539    &2007    &1018   &1425  \\
\hline
\end{array}
}
{\bf Neutral CP-odd  Higgs}
\eqn{
\begin{array}{|c|c|c|c|c|c|c|c|c|c|}
\hline
i  &1      &2     &3     &4       &5       &6       &7     &8     &9  \\
\hline
u_i&-0.703 & 0.073&-0.040& 0.004  & 0.0005 & 0.700  & 0.074&-0.002&-0.059\\
   & 0.703 &-0.073&-0.040& 0.004  &-0.0005 & 0.700  & 0.074&-0.002&-0.059\\
   & 0     & 0    &-0.889& 0.086  & 0      &-0.088  &-0.009&-0.029&-0.044\\
   & 0.0014& 0.008& 0.008&-0.00008& 0.707  & 0.0003 & 0.008&-0.706& 0.030\\
   &-0.0014& 0.008& 0.008&-0.00008&-0.707  & 0.0003 & 0.008&-0.706& 0.030\\
   & 0     & 0    & 0.448&-0.004  & 0      &-0.049  &-0.005&-0.033&-0.892\\
   &-0.073 &-0.703&-0.006&-0.070  & 0.008  & 0.074  &-0.700&-0.008&-0.002\\
   & 0.073 & 0.703&-0.006&-0.070  &-0.008  & 0.074  &-0.700&-0.008&-0.002\\
   & 0     & 0    &-0.079&-0.991  & 0      &-0.012  & 0.099&-0.001&-0.034\\
\hline
m_i&112.12 &532   &0     &0       &2005    &145.53  &535   &2008  &1423  \\
\hline
\end{array}
}
{\bf Charged Higgs}
\eqn{
\begin{array}{|c|c|c|c|c|c|c|}
\hline
i  &1        &2        &3      &4     &5       &6     \\
\hline
u_i& 0.707   &-0.000055&-0.703 & 0.041&-0.001  & 0.059\\
   &-0.707   & 0.000055&-0.703 & 0.041&-0.001  & 0.059\\
   & 0       & 0       & 0.089 & 0.893&-0.029  & 0.441\\
   & 0.000055& 0.707   &-0.0018&-0.008&-0.706  &-0.030\\
   &-0.000055&-0.707   &-0.0018&-0.008&-0.706  &-0.030\\
   & 0       & 0       & 0.049 &-0.447&-0.032  & 0.893\\
\hline
m_i&90.70    &2005     &130.02 &0     &2007    &1422  \\
\hline
\end{array}
}
{\bf Neutralino}\\
For neutralinos, the diagonalization matrix is defined as
\eqn{
&&U^TMU=\mbox{diag}(m_1,m_2,\cdots m_{12}). \no \\
&&
\begin{array}{|c|c|c|c|c|c|c|}
\hline
i   &1     &2     &3     &4     &5     &6\\
\hline
u_i & 0.234&-0.496& 0.446&-0.248&-0.475& 0.067  \\
    &-0.234& 0.496&-0.446&-0.248&-0.475& 0.067  \\
    &0     & 0    & 0    & 0.064& 0.148& 0.031  \\
    & 0.209& 0.504& 0.450&-0.217& 0.497&-0.055  \\
    &-0.209&-0.504&-0.450&-0.217& 0.497&-0.055  \\
    & 0    & 0    & 0    & 0.028&-0.105&-0.066  \\
    &-0.634&-0.017& 0.313& 0.621&-0.023&-0.002  \\
    & 0.634& 0.017&-0.313& 0.621&-0.023&-0.002  \\
    & 0    & 0    & 0    &-0.086& 0.005&-0.00007\\
    & 0    & 0    & 0    & 0.009& 0.084& 0.965  \\
    & 0    & 0    & 0    &-0.014&-0.119&-0.221  \\
    & 0    & 0    & 0    &0.0004& 0.001& 0.003  \\
\hline
m_i &-41.92&-130  &172   &-44.55&-111  &-177    \\
\hline
\hline
i  &7      &8      & 9    &10    &11    &12\\
\hline
u_i& 0.062 &-0.430 &-0.099&-0.099&-0.020& 0.002\\
   & 0.062 &-0.430 &-0.099&-0.099&-0.020& 0.002\\
   & 0.039 & 0.132 &-0.689&-0.682&-0.121&-0.003\\
   &-0.049 &-0.439 & 0.062&-0.063&-0.013&-0.003\\
   &-0.049 &-0.439 & 0.062&-0.063&-0.013&-0.003\\
   &-0.077 & 0.085 & 0.692&-0.686&-0.138&-0.050\\
   &-0.002 &-0.330 & 0.001&-0.003&-0.053& 0.047\\
   &-0.002 &-0.330 & 0.001&-0.003&-0.053& 0.047\\
   &-0.0001& 0.051 & 0.014& 0.114&-0.728& 0.669\\
   &-0.241 & 0.007 & 0.063&-0.014&-0.002&-0.001\\
   &-0.960 &-0.012 &-0.118& 0.024& 0.003& 0.002\\
   & 0.003 &-0.0009&-0.032& 0.153&-0.655&-0.739\\
\hline
m_i&-194   &157    &832   &820   &945   &-1144 \\
\hline
\end{array}
}
{\bf Chargino}\\
For the charginos, the diagonalization matrices are defined as
\eqn{
&&\chi^-=U_-(\chi^-)',\quad \chi^+=U_+(\chi^+)' , \no \\
&&U^T_+MU_-=diag(m_1,m_2,m_3,m_4)=(130,111,193,830), \no \\
&&U_+=\L(
\begin{array}{cccc}
 0.707 & 0.685 & 0.104 & 0.140 \\
-0.707 & 0.685 & 0.104 & 0.140 \\
 0     &-0.209 & 0.066 & 0.976 \\
 0     & 0.131 &-0.987 & 0.095 \\
\end{array}
\R),\quad
U_-=\L(
\begin{array}{cccc}
-0.707 &-0.696 & 0.085 & 0.088 \\
 0.707 &-0.696 & 0.085 & 0.088 \\
 0     & 0.140 & 0.128 & 0.982 \\
 0     &-0.102 &-0.984 & 0.143 \\
\end{array}
\R).\no \\
}
These mass eigenvalues are consistent with the experimental mass bounds \cite{PDG2008}
\eqn{
\mbox{Charged Higgs}&:&m \geq 79.3\mbox{GeV} ,\no \\
\mbox{Neutral CP-even Higgs}&:&m \geq 114.4\mbox{GeV} ,\no \\
\mbox{Neutral CP-odd Higgs}&:&m \geq 93.4\mbox{GeV} ,\no \\
\mbox{Chargino}&:&m \geq 94\mbox{GeV} \no ,\\
\mbox{Neutralino}&:&m \geq 46\mbox{GeV} \no .
}
Note that $\rho'_3,\rho'_4,(H^\pm_4)'$ are Nambu-Goldstone boson which are eaten by
gauge bosons.

\section{The lifetimes of exotic quarks}
As the R-parities of exotic quarks are odd, at least there must be one sfermion which is
lighter than exotic quarks, to make them unstable. Now we assume the right handed slepton $E^c_1$
is lighter than $2TeV$ and the other sfermions are heavier than $4TeV$. 
For simplicity, we assume there is no mixing between $E^c_1$ and $L_i$.
Through the non-renormalizable interaction
\eqn{
{\cal L} \supset -\sum_i\frac{c_iV^2}{\sqrt{3}M^2_P}(\psi_{g,1}+\psi_{g,2}+\psi_{g,3})E^c_1u^c_i+h.c. ,
}
the exotic quarks $\psi_{g,1\sim 3}$ can decay into $u^c_i$ and $E^c_1$, where $c_i$ are $O(1)$ coefficients. The lifetimes are estimated as follows
\eqn{
\Gamma(\psi^\dagger_{g,j}\to E^c_1+u^c_i)=\frac{2TeV}{16\pi}\L(\frac{c_iV^2}{\sqrt{3}M^2_P}\R)^2
=\frac{c^2_i}{1.7[\mbox{sec}]},
}
from which we must put $c_i\sim 4$ 
in order to satisfy the cosmological constraint for exotic particle, $\tau <0.1 \mbox{sec}$.
The interaction Eq.(132) comes from
\eqn{
W \supset \sum_i\frac{c_i}{M^2_P}\Phi_3(\Phi^c_1g_1+\Phi^c_2g_2+\Phi^c_3g_3)E^c_1U^c_i,
}
which may contribute to LF decay through
\eqn{
{\cal L} \supset -\sum_{i,j}\frac{c_iV}{M^2_P}\alpha_j \phi_{LF}\psi_{g,j}E^c_1u^c_i+h.c.,
}
where $\alpha_j$ are given by linear combinations of the  flavon mixing parameters.
In this paper, we assume $\alpha_j$ are small enough and this interaction does not
give sizable contribution to LF decay width.

\section*{Acknowledgments}
This work is supported by Young Researcher Overseas Visits Program for Vitalizing Brain Circulation Japanese in JSPS (T.T.). 
H.O.\ acknowledges
partial supports from the Science and Technology Development Fund
(STDF) project ID 437 and the ICTP project ID 30.



\end{document}